\documentclass[11pt,a4paper]{article}
\usepackage{amssymb}
\usepackage{amscd}
\usepackage[matrix,arrow,curve]{xy}
\begin{document}


\newcounter{mo}
\newcommand{\mo}[1]
{\stepcounter{mo}$^{\bf MO\themo}$%
\footnotetext{\hspace{-3.7mm}$^{\blacksquare\!\blacksquare}$
{\bf MO\themo:~}#1}}

\newcounter{ale}
\newcommand{\ale}[1]
{\stepcounter{ale}$^{\bf ALE\theale}$%
\footnotetext{\hspace{-3.7mm}$^{\blacksquare\!\blacksquare}$ {\bf
ALE\theale:~}#1}}

\newcounter{az}
\newcommand{\az}[1]
{\stepcounter{az}$^{\bf AZ\theaz}$%
\footnotetext{\hspace{-3.7mm}$^{\blacksquare\!\blacksquare}$ {\bf
AZ\theaz:~}#1}}


\newcommand{\Si}{\Sigma}
\newcommand{\tr}{{\rm tr}}
\newcommand{\ad}{{\rm ad}}
\newcommand{\Ad}{{\rm Ad}}
\newcommand{\ti}[1]{\tilde{#1}}
\newcommand{\om}{\omega}
\newcommand{\Om}{\Omega}
\newcommand{\de}{\delta}
\newcommand{\al}{\alpha}
\newcommand{\te}{\theta}
\newcommand{\vth}{\vartheta}
\newcommand{\be}{\beta}
\newcommand{\la}{\lambda}
\newcommand{\La}{\Lambda}
\newcommand{\D}{\Delta}
\newcommand{\ve}{\varepsilon}
\newcommand{\ep}{\epsilon}
\newcommand{\vf}{\varphi}
\newcommand{\vfh}{\varphi^\hbar}
\newcommand{\vfe}{\varphi^\eta}
\newcommand{\fh}{\phi^\hbar}
\newcommand{\fe}{\phi^\eta}
\newcommand{\G}{\Gamma}
\newcommand{\ka}{\kappa}
\newcommand{\ip}{\hat{\upsilon}}
\newcommand{\Ip}{\hat{\Upsilon}}
\newcommand{\ga}{\gamma}
\newcommand{\ze}{\zeta}
\newcommand{\si}{\sigma}

\def\hS{{\hat{S}}}

\newcommand{\li}{\lim_{n\rightarrow \infty}}
\def\mapright#1{\smash{
\mathop{\longrightarrow}\limits^{#1}}}

\newcommand{\mat}[4]{\left(\begin{array}{cc}{#1}&{#2}\\{#3}&{#4}
\end{array}\right)}
\newcommand{\thmat}[9]{\left(
\begin{array}{ccc}{#1}&{#2}&{#3}\\{#4}&{#5}&{#6}\\
{#7}&{#8}&{#9}
\end{array}\right)}
\newcommand{\beq}[1]{\begin{equation}\label{#1}}
\newcommand{\eq}{\end{equation}}
\newcommand{\beqn}[1]{\begin{eqnarray}\label{#1}}
\newcommand{\eqn}{\end{eqnarray}}
\newcommand{\p}{\partial}
\def\sq2{\sqrt{2}}
\newcommand{\di}{{\rm diag}}
\newcommand{\oh}{\frac{1}{2}}
\newcommand{\su}{{\bf su_2}}
\newcommand{\uo}{{\bf u_1}}
\newcommand{\SL}{{\rm SL}(2,{\mathbb C})}
\newcommand{\GLN}{{\rm GL}(N,{\mathbb C})}
\def\sln{{\rm sl}(N, {\mathbb C})}
\def\sl2{{\rm sl}(2, {\mathbb C})}
\def\SLN{{\rm SL}(N, {\mathbb C})}
\def\SLT{{\rm SL}(2, {\mathbb C})}
\def\PSLN{{\rm PSL}(N, {\mathbb C})}
\newcommand{\PGLN}{{\rm PGL}(N,{\mathbb C})}
\newcommand{\gln}{{\rm gl}(N, {\mathbb C})}
\newcommand{\PSL}{{\rm PSL}_2( {\mathbb Z})}
\def\f1#1{\frac{1}{#1}}
\def\lb{\lfloor}
\def\rb{\rfloor}
\def\sn{{\rm sn}}
\def\cn{{\rm cn}}
\def\dn{{\rm dn}}
\newcommand{\rar}{\rightarrow}
\newcommand{\upar}{\uparrow}
\newcommand{\sm}{\setminus}
\newcommand{\ms}{\mapsto}
\newcommand{\bp}{\bar{\partial}}
\newcommand{\bz}{\bar{z}}
\newcommand{\bw}{\bar{w}}
\newcommand{\bA}{\bar{A}}
\newcommand{\bG}{\bar{G}}
\newcommand{\bL}{\bar{L}}
\newcommand{\btau}{\bar{\tau}}

\newcommand{\tie}{\tilde{e}}
\newcommand{\tial}{\tilde{\alpha}}

\newcommand{\hhr}{\check{r}}
\newcommand{\ttr}{\tilde{r}}
\newcommand{\aar}{\acute{r}}
\newcommand{\hhe}{\check{E}}
\newcommand{\tte}{\tilde{E}}
\newcommand{\aae}{\acute{E}}

\newcommand{\Sh}{\hat{S}}
\newcommand{\vtb}{\theta_{2}}
\newcommand{\vtc}{\theta_{3}}
\newcommand{\vtd}{\theta_{4}}

\def\mC{{\mathbb C}}
\def\mZ{{\mathbb Z}}
\def\mR{{\mathbb R}}
\def\mN{{\mathbb N}}

\def\frak{\mathfrak}
\def\gg{{\frak g}}
\def\gJ{{\frak J}}
\def\gS{{\frak S}}
\def\gL{{\frak L}}
\def\gG{{\frak G}}
\def\gk{{\frak k}}
\def\gK{{\frak K}}
\def\gl{{\frak l}}
\def\gh{{\frak h}}
\def\gH{{\frak H}}
\def\gt{{\frak t}}
\def\gT{{\frak T}}
\def\gR{{\frak R}}

\def\baal{\bar{\al}}
\def\babe{\bar{\be}}

\def\bfa{{\bf a}}
\def\bfb{{\bf b}}
\def\bfc{{\bf c}}
\def\bfd{{\bf d}}
\def\bfe{{\bf e}}
\def\bff{{\bf f}}
\def\bfg{{\bf g}}
\def\bfm{{\bf m}}
\def\bfn{{\bf n}}
\def\bfp{{\bf p}}
\def\bfu{{\bf u}}
\def\bfv{{\bf v}}
\def\bfr{{\bf r}}
\def\bfs{{\bf s}}
\def\bft{{\bf t}}
\def\bfx{{\bf x}}
\def\bfy{{\bf y}}
\def\bfM{{\bf M}}
\def\bfR{{\bf R}}
\def\bfC{{\bf C}}
\def\bfP{{\bf P}}
\def\bfq{{\bf q}}
\def\bfS{{\bf S}}
\def\bfJ{{\bf J}}
\def\bfz{{\bf z}}
\def\bfnu{{\bf \nu}}
\def\bfsi{{\bf \sigma}}
\def\bfU{{\bf U}}
\def\bfso{{\bf so}}

\def\clA{\mathcal{A}}
\def\clC{\mathcal{C}}
\def\clD{\mathcal{D}}
\def\clE{\mathcal{E}}
\def\clG{\mathcal{G}}
\def\clR{\mathcal{R}}
\def\clU{\mathcal{U}}
\def\clT{\mathcal{T}}
\def\clO{\mathcal{O}}
\def\clH{\mathcal{H}}
\def\clK{\mathcal{K}}
\def\clJ{\mathcal{J}}
\def\clI{\mathcal{I}}
\def\clL{\mathcal{L}}
\def\clM{\mathcal{M}}
\def\clN{\mathcal{N}}
\def\clQ{\mathcal{Q}}
\def\clW{\mathcal{W}}
\def\clZ{\mathcal{Z}}

\def\baf{{\bf f_4}}
\def\bae{{\bf e_6}}
\def\ble{{\bf e_7}}
\def\bag2{{\bf g_2}}
\def\bas8{{\bf so(8)}}
\def\baso{{\bf so(n)}}

\def\sr2{\sqrt{2}}
\newcommand{\ran}{\rangle}
\newcommand{\lan}{\langle}
\def\f1#1{\frac{1}{#1}}
\def\lb{\lfloor}
\def\rb{\rfloor}
\newcommand{\slim}[2]{\sum\limits_{#1}^{#2}}

\newcommand{\sect}[1]{\setcounter{equation}{0}\section{#1}}
\renewcommand{\theequation}{\thesection.\arabic{equation}}
\newtheorem{predl}{Proposition}[section]
\newtheorem{defi}{Definition}[section]
\newtheorem{rem}{Remark}[section]
\newtheorem{cor}{Corollary}[section]
\newtheorem{lem}{Lemma}[section]
\newtheorem{theor}{Theorem}[section]

\begin{flushright}
 ITEP-TH-35/12\\
\end{flushright}
\vspace{15mm}
\begin{center}
{\Large{\bf Characteristic Classes of $\SLN$-Bundles and \\ \vspace{2mm} Quantum Dynamical Elliptic R-Matrices}
}\\
\vspace{8mm}
 {\bf A. Levin},$^{\natural\ \sharp}$ {\bf M. Olshanetsky},$^{\sharp}$ {\bf A. Smirnov},$^{ \flat\ \sharp}$
 {\bf A. Zotov}$^{\ \sharp}$\\ \vspace{5mm}
 \vspace{3mm} $^\sharp$ - {\sf Institute of Theoretical and Experimental Physics, Moscow, 117218, Russia}\\
 \vspace{2mm}$^\natural$ - {\sf Laboratory of Algebraic Geometry, GU-HSE, 7 Vavilova Str., Moscow, 117312, Russia}\\
 \vspace{2mm} $^\flat$ - {\sf Math.Dept., Columbia University, New York, NY 10027, USA}\\
 \vspace{4mm} E-mails: {\em alevin57@gmail.com}; {\em olshanet@itep.ru}; {\em asmirnov@itep.ru}; {\em zotov@itep.ru}\\
 \vspace{5mm}
\end{center}

\vspace{4mm}

\begin{abstract}
We discuss quantum dynamical elliptic R-matrices related to
arbitrary complex simple Lie group G.
 They  generalize the known vertex  and dynamical R-matrices and play an  intermediate role between these two types.
 The   R-matrices are defined by the corresponding characteristic classes describing the underlying vector bundles.
  The latter are related to elements of the center $\clZ(G)$ of G.
 While the known dynamical R-matrices are related to the bundles with trivial characteristic classes,
 the Baxter-Belavin-Drinfeld-Sklyanin vertex R-matrix corresponds to the  generator of the center $Z_N$ of SL(N).
 We construct    the R-matrices related
 to SL(N)-bundles with an arbitrary characteristic class explicitly and discuss the corresponding IRF models.
\end{abstract}

\vfill\eject

\tableofcontents


\section{Introduction}
\setcounter{equation}{0}

The quantum dynamical $R$-matrices and the Quantum Dynamical Yang-Baxter (QDYB) Equation,
 they satisfied,  introduced by G.Felder \cite{Fe1,Fe}, while
without the spectral parameter these structures appeared earlier \cite{GNe,Ba,AF}. The classical version of the QDYB equation
is the Classical Dynamical Yang-Baxter Equation and its solution is the classical  dynamical $r$-matrix. The classification
of the  classical  dynamical $r$-matrices with the elliptic spectral parameter was proposed in \cite{EV1}.

 The standard elliptic
quantum $R$-matrix does not depend on dynamical variables \cite{Baxter,Belavin} (see also \cite{KRS,BD}).
 It is defined only for the group $G=\SLN$. We will refer to it as Baxter-Belavin-Drinfeld-Sklyanin $R$-matrix.
 The quantum $R$-matrix  and the Yang-Baxter equation are the key
tools for the Quantum Inverse Scattering Method \cite{Skl2}-\cite{KuRe}. In particular, they define the commutation relations
in the vertex-type models and the corresponding Sklyanin-type algebras \cite{Skl1,FO,CLOZ2}.

On the other hand, the Felder
$R$-matrix depends on additional dynamical variable $\bfu\in\gh$, where $\gh$ is a Cartan subalgebra
of $\gg=\hbox{Lie}\,(G)$. It  is related to the IRF models of statistical mechanics \cite{Fe}.

Here we consider an intermediate situation. It  arises when $G$ is a  simple Lie group with a non-trivial centers,
 i.e. when $G$ is a classical group or $E_6$ and $E_7$. In these cases
 the dynamical parameter belongs to
 some subalgebra $\ti\gh^0$ of the Cartan subalgebra $\gh$ of the Lie algebra $\gg$.
At the classical level the problem was
 investigated in our previous papers \cite{LZ,LOSZ1,LOSZ2,LOSZ3}. In particular, we constructed there
  the classical elliptic $r$-matrices. They  complete the  Etingof-Varchenko classification \cite{EV1},
  where the case  $\ti\gh^0=\gh$  was considered.

   The quantum version of classical construction  \cite{EV1} was proposed in  \cite{EV2}.
The construction of the quantum $R$-matrices is more elaborate. It depends essentially on the representation
space $V$.  We suggest that upon the appropriate choice of $V$ the $R$-matrices can be constructed
in the general situation as in the case   $\ti\gh^0=\gh$ and $G=\SLN$.
 We demonstrate it explicitly $R$-matrices for $\SLN$ with elliptic dependence of the spectral parameter.
If $N$ is a prime number there is only two types of $R$-matrices exists. The first one is the
Baxter-Belavin-Drinfeld-Sklyanin vertex $R$-matrix \cite{Ba,Belavin}, and the second type is the Felder dynamical $R$-matrix
\cite{Fe}. But if $N=pl$, $(p\neq 1,N)$, then there exists new types of $R$-matrices. Their construction is the main result
of this paper.

While different universal structures related to the Yang-Baxter equations are well studied
for arbitrary simple Lie group
in trigonometric and rational cases \cite{yang1}-\cite{yang11}, the elliptic solution of the QDYB equation
 with spectral
parameter (\ref{qdyb}) is known only in $\SLN$ case \cite{ell_qg1}-\cite{ell_qg6}.


In the $A_{N-1}$ case the center of $G\!=\!\SLN$ is the cyclic group $\mu_N=\mZ/N\mZ$.
Represent elements of $\mu_N$ as $\exp\,\frac{2\pi i}Nj\,,~j=0\ldots,N-1$.
Then the Felder's case corresponds to $j=0$ while the Baxter-Belavin-Drinfeld-Sklyanin's one appears from $j=1$.
The intermediate situation takes place when $j=p>1$ and $N=pl$. In this case $\dim\ti\gh={\hbox{g.c.d}}(j,N)=p>1$.

{\bf The purpose of the paper} is to construct the quantum elliptic dynamical $R$-matrix in the intermediate case. The answer
is given by the Theorem \ref{theor} (Section \ref{sec4}).
 It is shown that the suggested $R$-matrix satisfies the QDYB equation
(\ref{qdyb}) with $\bfu\in C\subset\ti\gh_0$ (ref{C}).
 The result is schematically presented in Table 1. The last column is the case of our
interest.

\bigskip

\begin{tabular}{|c|c|c|c|}
  \hline
  &&&\\
  $\zeta=$ & 1 &$\exp\,(\frac{2\pi i}N)$ & $\exp\,(\frac{2\pi i}Np)\,,$\,~~$N=pl$ \\
   \hline
   &&&\\
 $R$-matrix & Felder Case & Baxter-Belavin-Drinfeld-Sklyanin Case & Intermediate Case \\
  \hline
\end{tabular}
\begin{center}
\textbf{Table 1.} \\
$R$-matrices corresponding to different characteristic classes of SL$(N)$ bundles.
\end{center}
\bigskip

The  classical integrable system corresponding to the intermediate case
 is the system of interacting elliptic tops \cite{LZ}. Our goal is to quantize its
classical $r$-matrix.

It should be mentioned that the dynamical and non-dynamical elliptic $R$-matrices are related by the dynamical twist
\cite{Pa,Has}( see also \cite{twist1}-\cite{twist4}). This twist was interpreted as a modification of bundle (or Hecke
transformation) in \cite{LOZ1}.
 At the classical level it acts by a singular gauge transformation on the Lax matrices and relates the models of
Calogero-Ruijsenaars type \cite{RS}-\cite{KKS} with the elliptic Euler-Arnold tops \cite{RSTS,LOZ1,KLO}.
In the theory of integrable models of statistical mechanics this Hecke transformation defines
 a twist providing a  passage from the so-called IRF type models \cite{Baxter3,JMO} to the Vertex type models
  \cite{Pa,Has}, \cite{irf-vertex1}-\cite{irf-vertex4}.
 In the isomonodromic deformations problem \cite{LO,Ta} corresponding to the Hitchin systems \cite{H}-\cite{ER}
on elliptic curves the modification relates Painlev\'e VI equation and nonautonomous Zhukovsky-Volterra gyrostat
 \cite{LOZ3,OZ_Jap,CLOZ1,LZ2}. The field (1+1) generalizations of the Hitchin-Nekrasov (Gaudin)
models are discussed in \cite{LOZ1,Kr2,Z}. They describe the continuous limit of the integrable chains.
In terms of a gauge field theory the Hecke transformation can be explained as a monopole solution of the Bogomolny equation
\cite{KW,LOZ2,Bul}.

The paper is organized as follows: in  the next section we review construction of bundles over elliptic curves and define
classical and quantum elliptic $R$-matrices, related to these bundles.
 In Section 3 we, first, recall the known quantum $R$-matrices corresponding to the first
and second columns in the Table 1.
 Then the quantum $R$-matrix for the intermediate case is suggested (\ref{x31}) and the QDYB equation is
verified (Theorem 3.1). Finally, we discuss possible applications of the obtained solution of the QDYB equation to IRF
models.

\vspace{0.3cm}
\bigskip
{\small {\bf Acknowledgments.}\\
 The work was supported by grants RFBR-09-02-00393, RFBR-09-01-93106-NCNILa (A.Z. and A.S.),
and by the Federal Agency for Science and Innovations of Russian Federation under contract 14.740.11.0347.
The work of A.Z. was also supported by the Dynasty fund and the President fund MK-1646.2011.1. The work of A.L. was partially
supported by AG Laboratory GU-HSE, RF government grant, ag. 11 11.G34.31.0023.}


\section{Characteristic Classes of Bundles over Elliptic Curves and $R$-Matrices}
\setcounter{equation}{0}

\subsection{Characteristic classes of bundles over elliptic curves}

 Let $G$ be a complex simple Lie group with a
nontrivial center $\clZ(G)$.
 A  simply-connected group $\bar G$
 in all cases apart $G_2$, $F_4$ and $E_8$ has a non-trivial center
 $\clZ(\bar G)\sim P^\vee/Q^\vee$, where $P^\vee$, ($Q^\vee$) is a coweight (coroot) lattice in the Cartan subalgebra
 $\gh$.  The center $\clZ(\bG)$ is a cyclic group except $\gg=D_{2n}$. In the latter case
 the group $\bar G=Spin_{4n}(\mC)$ has a non-trivial center
$\clZ(Spin_{4n})=(\mu^L_2\times\mu^R_2)\,,~~\mu_2=\mZ/2\mZ$.

  The adjoint group is the quotient $G^{ad}=\bar G/\clZ(\bar G)$.
For the cases $A_{n-1}$  (when $n=pl$ is non-prime) and $D_n$ the center $\clZ(\bar G)$ has non-trivial subgroups
$\clZ_l\sim\mu_l=\mZ/l\mZ$. Assume that $(p,l)$ are co-prime.   There exists the quotient-groups
 \beq{fgl}
G_l=\bG/\clZ_l\,,~~~G_p=G_l/\clZ_p\,,~~~G^{ad}=G_l/\clZ(G_l)\,,
 \eq
where $\clZ(G_l)$ is the center of $G_l$ and $\clZ(G_l)\sim\mu_p=\clZ(\bar G)/\clZ_l$. Let $N=\hbox{Ord}(\clZ(\bar G))$. Then
we come to the diagram:
$$
\xymatrix{
&&G^{ad}&\\
&G_p\ar[ur]&G^{ad}&G^{ad}\\
\mu_l\ar[ur]\ar[r]&\bG\ar[ur]\ar[u]\ar[r]&G_l\ar[ur]&\\
\mu_N\ar[ur]&\mu_p\ar[u]\ar[ur]&&
}
$$

Let $\clE_G$ is a principle $G$-bundle  over an elliptic curve $\Si_\tau=\mC/(\mZ+\tau\mZ)$. We define a holomorphic
$G$-bundle  $E=\clE_G\times_GV$ (or simply $E_G$) over $\Si_\tau$. The bundle $E_G$ has the space of sections
$\G(E_G)=\{s\}$, where $s$ takes values in $V$. The bundle $E_G$ is defined by transition matrices of its sections around the
fundamental cycles. Then sections of $E_G(V)$  assume the quasi-periodicities:
  \beq{mon1}
s(z+1)=\clQ(z)\,s(z)\,,~~s(z+\tau)=\La(z)\, s(z)\,,
 \eq
where $\clQ(z)\,,\La(z)$ take values in the representation of $G$. Then
  $\clQ(z)\,,\La(z)$ satisfy the following equation
\beq{1eq}
\clQ(z+\tau)\La_j(z)\clQ(z)^{-1}\La_j^{-1}(z+1)=Id\,.
\eq
 It  follows from  \cite{NS} that it is possible to choose the constant
 transition operators.  Then we come to the equation
  \beq{1}
  \clQ\La\clQ^{-1}\La^{-1}=Id\,.
  \eq
    Solutions of this equation are defined up to conjugations form
  the moduli space of $E_{\bG}$ bundles over $\Si_\tau$. We can modify (\ref{1}) as
  $$
    \clQ\La\clQ^{-1}\La^{-1}= \zeta Id\,,
 $$
 where $\zeta $ is a generator of the center $\clZ(\bG)$. In this case $(\clQ,\La)$ are the clutching operators
 for $G^{ad}$-bundles, but not for $\bG$-bundles, and $\zeta$ plays the role of obstruction to lift the $G^{ad}$-bundle
 to the $\bG$-bundle.

 More generally, consider solutions of the equation
 \beq{o01}
  \clQ\La^p\clQ^{-1}\La_j^{-p}= \zeta^p Id\,,
 \eq
 where $p=\hbox{ord}(\clZ(\bG))/l$, and therefore $\zeta^p \in\clZ_l$. It means that  $\zeta^p$ is obstruction to
 lift the $G_l$-bundle to  the $\bG$-bundle.

The obstructions can be formulated in terms of cohomology of $\Si_\tau$. Namely,
the first cohomology $H^1(\Si_\tau,G(\clO_{\Si_\tau}))$  of $\Si_\tau$ with coefficients
in analytic sheaves $G(\clO_{\Si_\tau})$ define  the moduli space $\clM(G,\Si_\tau)$
of holomorphic $G$-bundles over $\Si_\tau$. Using  (\ref{fgl}) we write three exact sequences
$$
\begin{array}{l}
  1\to\clZ(\bG))\to\bG(\clO_{\Si_\tau})\to G^{ad}(\clO_{\Si_\tau})\to 1\,, \\
    1\to\clZ_l\to\bG(\clO_{\Si_\tau})\to G_l(\clO_{\Si_\tau})\to 1\,, \\
    1\to\clZ(G_l)\to G_l(\clO_{\Si_\tau})\to G^{ad}(\clO_{\Si_\tau})\to 1\,.
\end{array}
$$
Then we come to the long exact sequences
\beq{coh1}
\to H^1(\Si_\tau,\bG(\clO_\Si))\to H^1(\Si_\tau,G^{ad}(\clO_\Si))\to H^2(\Si_\tau,\clZ(\bG))\sim\clZ(\bG))\to 0 \,,
\eq
\beq{coh2}
\to H^1(\Si_\tau,\bG(\clO_\Si))\to H^1(\Si_\tau,G_l(\clO_\Si))\to H^2(\Si_\tau,\clZ_l)\sim\mu_l\to 0\,,
\eq
\beq{coh3}
\to H^1(\Si_\tau,G_l(\clO_\Si)\to H^1(\Si_\tau,G^{ad}(\clO_\Si))\to H^2(\Si_\tau,\clZ(G_l))\sim\mu_p\to 0\,.
\eq
The elements from $H^2$ are obstructions to lift bundles, namely
$$
\begin{array}{c}
\zeta(E_{G^{ad}})\in H^2(\Si_\tau,\clZ(\bG)) -{\rm ~obstructions ~to~ lift~}
 E_{G^{ad}}-{\rm bundle~to~}E_{\bG}-{\rm bundle}\,,\\
  \zeta(E_{G_l})\in H^2(\Si_\tau,\clZ_l) -{\rm ~obstructions ~to~ lift~}
 E_{G_l}-{\rm bundle~to~}E_{\bG}-{\rm bundle}\,, \\
    \zeta^\vee(E_{G^{ad}})\in H^2(\Si_\tau,\clZ(G_l)) -{\rm ~obstructions ~to~ lift~}
 E_{G^{ad}}-{\rm bundle~to~}E_{G^l}-{\rm bundle}\,.
  \end{array}
$$
\begin{defi}
Images of $H^1(\Si_\tau,G(\clO_\Si))$ in $H^2(\Si_\tau,\clZ)$ are called the characteristic classes
$\zeta(E_G)$ of $G$-bundles.
\end{defi}

Let $P^\vee$, ($Q^\vee$) be the coweight (the coroot) lattice in $\gh$. Then  $\clZ(\bG)=P^\vee/Q^\vee$
\cite{Bou}. If $\clZ(\bG)$ is cyclic,
\footnote{In the case $\bG=\hbox{Spin}_4n$ the center $\clZ(\hbox{Spin}_4n)$ is generated by
the coweights corresponding to the left and to the right spinors.}
then there exists a fundamental coweight $\varpi^\vee\in P^\vee$ generating $\clZ(\bG)$.
It means that $\hbox{Ord}(\clZ(\bG))\varpi^\vee\in Q^\vee$. Similarly, $\clZ_l$ is generated by the coweight
$\varpi_j^\vee$ such that $l\varpi_j^\vee\in Q^\vee$.

It was proved in \cite{LOSZ1} that for generic bundles the solution of the equation (\ref{o01}) can be written as
\footnote{For brevity we omit the index $j$ in (\ref{o01}).}
 \beq{qala}
\clQ=\bfe\,(\ka)\,,~~~~\La^p=\bfe\,(\bfu)\La_0\,.
 \eq
 $$
\ka=\bfe\,(\rho^\vee/h)\,,~~~\rho^\vee=\oh\sum_{\al^\vee>0}\al^\vee\,,~~h-{\rm Coxeter~number}\,,
 $$
where $\{\al^\vee\}$ are co-roots of $\gg$.
The element $\La_0$ is uniquely defined by the element from $ \clZ(G)$ which is an element of the Weyl group $W(\gh)$ that
acts as a symmetry of the extended Dynkin diagram \cite{Bou}. Moreover,  $\bfu\in Ker\,(\La_0-Id)$ and  $\bfu$ belongs to
Cartan subalgebra $\ti\gh_0\subset\gh$, where $\gh$ is a Cartan subalgebra containing $\clQ$.
 An element from  $\ti\gh_0$ plays the role of a parameter in the moduli space $\clM_G$
of holomorphic bundles $E_G$ over $\Si_\tau$. The subalgebra   $\ti\gh_0$ is a Cartan subalgebra of invariant
subgroup $\ti\gg_0$ \cite{LOSZ1}.

More exactly, $\bfu$ belongs to a fundamental domain $C^\G$ of the affine Weyl group $W\ltimes(\tau \G^\vee\oplus \G)$, where
$\G$ is a sublattice in  $\ti\gh_0$ $Q_l^\vee\subseteq\G\subseteq P_l^\vee$:
 \beq{C}
 \bfu\in C^\G=\ti\gh_0/\Bigl(W\ltimes(\tau \G\oplus \G)\Bigr)\,.
 \eq
The subalgebras $\ti\gh_0\subset\gh$ were classified in \cite{LOSZ1}.


\subsection{Elliptic $R$-Matrices}

A general form of the quantum dynamical (modified) Yang-Baxter (QDYB) equation related to a simple complex Lie group $G$ has
the following form. Let $\gh$ be a commutative subalgebra $\gh\subset \gg=$Lie$\,G$, $\gh^*$ is the dual space. Consider
finite-dimensional $G$-modulus  $V_j$, $(j=1,2,3)$  and let  $V=\oplus_{\mu\in P}V_\mu$. be the weight decomposition. Let
$z\in\mC$ be the spectral parameter. The quantum elliptic  dynamical $R$-matrix is the map $\gh^*\times\mC\to
\hbox{Aut}(V_j\otimes V_k)$, $(j,k=1,2,3)$, $j\neq k$, depending on the Plank constant $\hbar$. $R$ satisfies the following
conditions:

\begin{itemize}
 \item $R$ has fixed periodicities with respect to the lattice $\mZ+\tau\mZ\subset\mC$.
Let  $\clQ$ and $\La$ be some fixed elements of $G$ (transition functions). Then
 \beq{qp}
 R(\bfu,z+1)=\hbox{Ad}_\clQ R(\bfu,z)\,,~~R(\bfu,z+\tau)=\hbox{Ad}_{\La^p(\bfu)} R(\bfu,z)\,,
 \eq
 where the adjoint operators (\ref{qala}) act on the first factor $V_j\otimes V_k$.
 It means that   $R(\bfu,z)$ is a section of a bundle  $\hbox{Aut}(V_j\otimes V_k)$ over the elliptic curve
 $\Si_\tau=\mC/(\mZ\oplus\tau\mZ)$ and the dynamical parameter $\bfu$ play the role of the tangent vector to
 the moduli space of the bundle.
                                                      \item
 $R$  satisfies the QDYB equation in $\hbox{End}\,(V\otimes V\otimes V)$
 \beq{qdyb}
R^{12}(\bfu-\hbar e^3,z_{12})R^{13}(\bfu+\hbar e^2,z_{13})R^{23}(\bfu-\hbar e^1,z_{23})=
 \eq
 $$
R^{23}(\bfu+\hbar e^1,z_{23})R^{13}(\bfu+\hbar e^2,z_{13})R^{12}(\bfu,
+\hbar e^3,z_{12})\,,~~~(z_{jk}=z_j-z_k)\,.
 $$
 The shift of the dynamical parameter $\bfu-\hbar e^j$ means, for example, that
 $R_{V_1V_2}^{12}(\bfu-\hbar e^3)$  acts on the tensor product $v_1\otimes v_2\otimes v_3$ as
 $R_{V_1V_2}^{12}(\bfu-\hbar \mu_3)$ if $v_3\in V_3[\mu_3]$.

                                                      \item The unitarity condition
 \beq{uc}
R^{12}(\bfu,z_{12})R^{21}(\bfu,z_{21})=Id_{V\otimes V}\,,~
 \eq
                                                      \item The weight zero condition
 \beq{z}
[X^1+X^2,R^{12}(\bfu,z_{12})]=0\,,~~~~\forall X\in\ti{\gh}_0\,.
 \eq
 \item The  quasi-classical limit $\hbar\to 0$
 \beq{qcl}
R(\bfu,z)=\frac{1}{\hbar}Id\otimes Id+r(\bfu,z)+O(\hbar)\,,
 \eq
where  $r(\bfu,z)$ is the classical dynamical $r$-matrix, related to the $E_G$ bundle, defined below. In this sense
$R(\bfu,z)$ is a quantization of the classical  dynamical $r$-matrix  $r(\bfu,z)$. In particular cases we obtain in this way
the classical non-dynamical elliptic Belavin-Drinfeld $r$-matrix \cite{BD}, or classical dynamical elliptic $r$-matrix
\cite{Skl4,BAB}. The latter type of $r$-matrices were classified in \cite{EV1}.
                                                    \end{itemize}
These conditions do not define  the $R$-matrix  uniquely. There are additional transformations corresponding
to shifts along the dynamical parameter. We will not discuss this issue here.

Let us focus now on the classical $r$-matrix in (\ref{qcl}). We define the classical  dynamical $r$-matrices
 (CDRM) following \cite{LOSZ1}.
In general case they
have the following form. To define the $r$-matrices for arbitrary characteristic classes let us define
the special basis (the
general sine basis (GS basis)) in the Lie algebra $\gg$ \cite{LOSZ1}.
 In the Cartan subalgebra it  is defined in the following way. Let $(e_1,e_2,\ldots,e_n)$ be a canonical basis in $\gh$,
 $((e_j,e_k)=\de_{jk})$
\footnote{For $A_n$ and $E_6$
root systems it is convenient to choose canonical bases in $\gh\oplus\mC$.}.
 Consider the $\La$ action ($\La^l=Id$). Since
$\La$ preserves $\gh$ we can consider the action $\la$ of $\La$ on the canonical basis.
Define an orbit  of length $l_s=l/p_s$
passing through $e_s$
 $$
\clO(s)=\{e_s,\la(e_{s}),\ldots,\la^{(l-1)}e_{s)}\}\,.
 $$
 $$
\gh_{s}^c=\f1{\sqrt{l}}\sum_{m=0}^{l-1}\om^{mc}\la^m(e_{s})\,,~~
c\in J_{p_s}\,, ~~\om=\exp\,(\frac{2\pi i}{l})\,,
 $$
where $J_{p_s}=\{c=mp_s\, mod(l)\,|\,m\in\mZ\}$.
Define the quotient $\clC_l=(e_1,e_2,\ldots,e_n)/\mu_l$.
Then we can pass from the canonical basis to the GS basis
 $$
(e_1,e_2,\ldots,e_n)\longleftrightarrow \{\gh_{s}^c\,,~s\in \clC_l\}\,.
 $$
Now replace the root basis $E_\be$ in $\gg$ by the GS-basis
 $$
\gt^a_{\babe}=\f1{\sqrt{l}}\,\sum_{m=0}^{l-1}\om^{ma}E_{\la^m(\be)}\,,~~
\om=\exp\,\frac{2\pi i}{l}\,,~~a\in J_\be\,.
 $$
This transformation is invertible $E_{\la^k(\be)}=\f1{\sqrt{l}}\sum_{a\in J_l}\om^{-ka}\gt^a_{\babe}$. The commutation
relations have the following form:
 \beqn{com1}
 [\gt^{a}_{\alpha},\gt^{b}_{\beta}]\,=\left\{
\begin{array}{ll}
\frac{1}{\sqrt{l}}\,\sum\limits_{s=0}^{l-1}\, \omega^{bs} \, C_{\alpha,\,
\lambda^s\beta}\,\gt^{a+b}_{\alpha+\lambda^s\beta},& \alpha\neq \,-\lambda^{s} \beta\,,\\ &
\\
\frac{p_{\alpha}}{\sqrt{l}}\,\omega^{s\,b}\,\gh^{a+b}_{\alpha}&\alpha= \,-\lambda^{s} \beta\,,
\end{array}\right.
  \eqn
\beqn{com2}
\begin{array}{l}
\left[\gh^{\,k}_{\,\alpha}, \gt^{\,m}_{\,\beta}\right] =
\frac{1}{\sqrt{l}}\,\sum\limits_{s=0}^{l-1}\,\omega^{-ks}\,\frac{2(\alpha,
\lambda^{s}\beta) }{(\alpha,\alpha)}\,\gt^{k+m}_{\,\beta}\,.\\
\end{array}
 \eqn
It is convenient to represent   $r(\bfu,z)$ as a sum of the Cartan and off-Cartan part:
 $$
 r(\bfu,z)=r(z)=r_{R}(z)+r_{\gh}(z)\,.
 $$
  \beq{Rma}
  r_{R}(z,w)=\frac{1}{2}\,\sum\limits_{a=\,0}^{l-1}\, \sum\limits_{\alpha\,\in\,R}\,
|\alpha|^2\,\varphi^{\,a}_{\,\alpha}(z-w) \,\gt^{\,a}_{\,\alpha}
\otimes \gt^{-a}_{-\alpha}\,,
 \eq
 $$
r_{\gh}(z,w)=
\sum\limits_{a=\,0}^{l-1}\,\sum_{\alpha\,\in\,\Pi}\,\varphi^{\,a}_{\,0}(z-w)\,
\gh^{\,a}_{\,\alpha}\otimes\gh^{-a}_{\,\alpha}\,,
 $$
 where functions $\varphi^{\,a}_{\,\alpha}(z)$ are defined in (\ref{kfi}).
%


\section{Quantum $R$-Matrices Related to $\SLN$}
\setcounter{equation}{0}

We apply the general construction of the quantum $R$-matrix to the case $G=\SLN$ and $V$ is the
standard vector representation. We pass from the GS-basis to the tensor basis (\ref{o04}) and write in
this basis the transition matrices $\clQ$ and $\La$.


\subsection{The Moduli Space of $\SLN$-Bundles over Elliptic Curves}

The dynamical parameter $\bfu$ belongs to  the moduli space of vector bundles over elliptic curves. We describe here. We
identify $\gh^*$ and $\gh\subset\sln$ by means of the standard metric on  $\mC^N$. The  roots and coroots for $\sln$ coincide
 and therefore the coroot lattice $Q^\vee$ coincides with the root lattice $Q$.
Let $\{e_j\}$ be the standard basis in $\mC^N$.
Then
\beq{rl}
Q=\{\sum m_je_j\,|\,m_j\in \mZ\,,~\sum m_j=0\}\,,
\eq
generated by the simple roots  $\Pi=\{\al_k\}$
$=\{\al_1=e_1-e_2\,\ldots,\al_{N-1}=e_{N-1}-e_N\}$.

The fundamental weights
$\varpi_k$, $(k=1,\ldots,N-1)$,  dual to the basis of simple
roots $\Pi^\vee\sim\Pi\,$ $\,(\varpi_k(\al^\vee_k)=\de_{kj})$, are
\beq{fw1}
\varpi_j=e_1+\ldots+e_j-\frac{j}{N}\sum_{l=1}^Ne_l\,,~~~
\left\{
\begin{array}{l}
 \varpi_1=(\frac{N-1}N,-\f1{N},\ldots,-\f1{N})\\
 \varpi_2=(\frac{N-2}N,\frac{N-2}N,\ldots,-\frac{2}{N})   \\
 \ldots \\
   \varpi_{N-1}=(\frac{1}N,\frac{1}N,\ldots,\frac{1-N}{N}) \,. \\
\end{array}
\right.
\eq

Similar to the roots and coroots we identify the
fundamental weights and the fundamental coweights. They generate the weight
(coweight) lattice
 \beq{wl}
 P\subset\gh\,,~P=\{\sum_ln_l\varpi_l\,|\,n_l\in\mZ\}\,,~~{\rm or}~
  P=\{\sum_{j=1}^Nm_je_j\,,~m_j\in \f1{N}\mZ\,,~m_j-m_k\in\mZ\}\,.
  \eq


The quotient-group  $ P/Q$ is isomorphic to the center $\clZ(\SLN)\sim\mu_N$. It is generated by $\zeta=\exp 2\pi
i\varpi_{1}$.

For the  \emph{ trivial bundles } corresponding to the Felder
$R$-matrix we have few moduli spaces $C^l$  (\ref{C}), corresponding
to a choice of the sublattice $P_l$. If $N$ is a prime number then
we have two options only . Let
$$
C^+=\{\bfu\in\mC^N\,|\,\Re e\,u_1\geq \Re e\,u_2\geq\ldots \geq \Re e\,u_N\}
$$
be a positive Weyl chamber . Then similar to the lattice $P_l$ in
(\ref{C}) we  may take  $Q$ $(l=1)$ (\ref{rl}), or $P$ $(l=N)$
(\ref{wl}). Then we come to the two types of alcoves
 \beq{csca}
C^{1} =\{\bfu\in C^+\,|\, u_j\sim u_j+\tau
n_j+m_j\,,~n_j,m_j\in\mZ\,,~\sum_j n_j=\sum_j m_j=0\}\,, \eq
 \beq{cada} C^{N} =\{\bfu\in C^{1}\,|\,n_j,m_j\in\f1{N}\mZ\,,~
n_j-n_k\in\mZ\,,~m_j-m_k\in\mZ~\}\,. \eq $C^{1}$ is the moduli space
of  $\SLN$-bundles, while $C^{N}$ is the moduli space of trivial
P$\SLN$-bundles (i.e  P$\SLN$-bundles that can be lifted to
$\SLN$-bundles).

If $N$ is not prime, then there are another sublattices of the weight lattice. For example, if $N=pl$ there are
$$
C^l=\{\bfu\in C^{1}\,|\,n_j,m_j\in\f1{l}\mZ\,,~
n_j-n_k\in\mZ\,,~m_j-m_k\in\mZ~\}\,,
$$
$$
C^p=\{\bfu\in C^{1}\,|\,n_j,m_j\in\f1{p}\mZ\,,~
n_j-n_k\in\mZ\,,~m_j-m_k\in\mZ~\}\,
$$
in addition to $C^1$ and $C^N$. They are the moduli space of trivial
$G_p=\SLN/\mu_p$ and $G_l=\SLN/\mu_l$ bundles. In general, the
number of different moduli spaces corresponds to the number of the
prime  factors of $N$.

Consider\emph{  nontrivial bundles} with transition matrices satisfying (\ref{o02}), where
$\zeta$ can be represented as $\zeta=\bfe(\varpi_p)$,
$$
\varpi_p=\Bigl(\underbrace{
\frac{N-p}N,\ldots\frac{N-p}N}_p,\underbrace{-\frac{p}N,\ldots-\frac{p}N}_{N-p}\Bigr)=
\Bigl(\frac{l-1}l,\ldots\frac{l-1}l,-\frac{1}l,\ldots-\frac{1}l\Bigr)\,,~~(l\varpi_p\in Q)\,.
$$
It is a generator of the group $\mu_l$, and in this way is the obstruction to lift the $G_l=\SLN/\mu_l$ bundle
to the $\SLN$-bundle (see (\ref{coh2})).
We consider the root $Q_l$ and the weight $P_l$ lattices in $\ti\gh_0$ (\ref{o03}).
 In the canonical basis $e_j=E_jj$, $(j=1,\ldots p)$ they have the form (\ref{rl}), (\ref{wl}).
In particular,
$$
P_l=\{\ga=\sum _{j=1}^pn_j\ti e_j\,,~~n_j\in\f1{p}\mZ\,,~~
\sum _{j=1}^pn_j=0\,,~~n_j-n_k\in\mZ\}\,.
$$
 It is an invariant sublattice of $P$.

 If $p$ is a prime number, then, similarly
 to (\ref{csca}) and (\ref{cada}) we have two types of the moduli spaces
 \beq{cl1}
C^{l,1}\,:\, u_j\sim u_j+\tau m_j+n_j\,,~n_j,m_j\in\mZ\,,~
\sum_{j=1}^p n_j=\sum_{j=1}^p m_j=0\,,
\eq
and for
\beq{clp}
C^{l, p}\,:\, u_j\sim u_j+\tau m_j+n_j\,,~~~~n_j,m_j\in\f1{p}\mZ\,,
\eq
$$
\sum_{j=1}^p n_j=\sum_{j=1}^p m_j=0\,,~~~n_j-n_k\in\mZ\,,~~~m_j-m_k\in\mZ\,.
$$
If $p$ is non-prime we have additional types of the moduli spaces as
it was given above for the trivial bundles.

\subsection{Description of the Adjoint Bundles and the Model of Interacting Tops}

In \cite{LZ} the classical integrable models corresponding to $\SLN$-bundles with nontrivial characteristic classes were
studied. Let us recall the results. We consider the $\SLN$-bundles with
 $$
N=lp\,,\ \  l,p\in{\mathbb Z}
 $$
defined by its multiplicators (\ref{mon1}) with the center element
 $$
\zeta=\exp(p\frac{2\pi i}{N})=\exp(\frac{2\pi i}{l})\in {\mathbb Z}/N{\mathbb Z}
 $$
from the condition (\ref{o01}). The multiplicators can be written explicitly in terms of $\SLN$-valued generators of the
finite Heisenberg group (\ref{q}) and (\ref{la}):
 \beq{o02}
\clQ\La^p\clQ^{-1}\La^{-p}=\exp(p\frac{2\pi i}{N}) Id\,,\ \ \ \clQ, \La\in\SLN\,.
 \eq
The dimension of the moduli space of these bundles equals g.c.d.$(N,p)=p$ \cite{Atiyah,Sch,FM}. Indeed, it is easy  to see
that the following Cartan element of the Lie algebra ${\bf u}\in \gh\subset\sln$ commutes with both $\clQ$ and $\La^p$:
 \beq{o03}
 \begin{array}{c}
{\bf u}=\hbox{diag}(u_1,...,u_p,u_1,...,u_p,...,u_1,...,u_p)=\bigoplus\limits_{j=1}^l {\bf u}_{p\times p}\,,
~~(\sum_{j=1}^pu_j=0)\,,\\
{\bf u}_{p\times p}=\hbox{diag}(u_1,...,u_p)\in\ti\gh_0\subset{{\rm sl}(p, {\mathbb C})}\,,\\ \ \\
\hbox{Ad}_{\clQ}{\bf u}=0,\ \ \ \hbox{Ad}_{\La}{\bf u}=0\,.
 \end{array}
 \eq
It was shown in \cite{LZ} that there exist such a number matrix  $S$ (combination of permutations) that
 \beq{vv3}
 \begin{array}{c}
S{\bf u}S^{-1}=\bigoplus\limits_{J=1}^p u_J Id_{l\times l}\,,\\
S\clQ S^{-1}=\bigoplus\limits_{J=1}^p\bfe\left(\frac{J-l}{N}\right)\clQ_{l\times l} \,,\\
S\Lambda^n S^{-1}=\bigoplus\limits_{J=1}^p \Lambda_{l\times l}\,.
 \end{array}
 \eq
The later means that any section $L(z)\in\Gamma(\hbox{End}(E_{\SLN}(V)))$
 has the following quasiperiodicity properties:
 \beq{vv5} \left\{
\begin{array}{l}
L_{IJ}(z+1)=\bfe\left(\frac{I-J}{N}\right) Q_{p\times p}L_{IJ}(z)Q_{p\times p}^{-1}
\\
L_{IJ}(z+\tau)=\bfe(-u_I)\Lambda_{p\times p}L_{IJ}(z)\Lambda_{p\times p}^{-1} \bfe(u_J)
\end{array}
\right.
 \eq
The factor $\bfe\left(\frac{I-J}{N}\right)$ can be removed by
$$
\begin{array}{c}
L_{IJ}(z)\rightarrow L_{IJ}(z)\bfe\left(-z\frac{I-J}{N}\right)
\\
u_I\rightarrow u_I-I\frac{\tau}{N}
\end{array}
$$
Finally, the boundary conditions are of the form:
 \beq{vv6} \left\{
\begin{array}{l}
L_{IJ}(z+1)= Q_{p\times p}L_{IJ}(z)Q_{p\times p}^{-1}
\\
L_{IJ}(z+\tau)=\bfe(-u_I)\Lambda_{p\times p}L_{IJ}(z)\Lambda_{p\times p}^{-1} \bfe(u_J)
\end{array}
\right.
 \eq
Therefore, it is natural to use the following basis
 \beq{o04}
  E^a_{ij}=E_{ij}\otimes T_a\in{{\rm gl}(N,{\mathbb C})}\,,\ \
   E_{ij}\in{{\rm gl}(p,{\mathbb C})}\,,\ \ T_a\in {{\rm gl}(l,{\mathbb
C})}
 \eq
in the Lie algebra ${{\rm gl}(N,{\mathbb C})}$, where $E_{ij}$ is a standard matrix basis in ${{\rm gl}(p,{\mathbb C})}$
(generated by the fundamental representation of ${{\rm GL}(p,{\mathbb C})}$) and $T_a$ is the basis of ${{\rm gl}(l,{\mathbb
C})}$ defined in (\ref{B.11})-(\ref{AA3b}). The basis (\ref{o04}) will be used in Section \ref{sec4} to construct the quantum
$R$-matrix.

The described type of bundles were used in \cite{LZ} in order to construct the "models of interacting tops".


\subsection{Baxter-Belavin-Drinfeld-Sklyanin and Felder Quantum $R$-matrices}

As we told there exist two extreme cases in the description of the $R$-matrices.
First case is the vertex $R$-matrices
\cite{Baxter,Belavin}. These $R$-matrices correspond the $\SLN$ bundles with the characteristic classes
$\zeta=\exp\,\frac{2\pi i k}N$, where $k$ and $N$ are coprime. In this case $\ti\gh_0=\emptyset$. It is so-called
non-dynamical $R$-matrices. Another case corresponds to the $\SLN$ bundles
 with the trivial  characteristic classes $\zeta=1$
\cite{Fe1,EV2}. We first consider the $R$-matrices for these two cases. The $R$-matrices satisfy the non-dynamical or
dynamical Yang-Baxter equations correspondingly. The later equations are
 \beq{z0}
R_{12}(z-w)R_{13}(z)R_{23}(w)= R_{23}(w)R_{13}(z)R_{12}(z-w)\,,
 \eq
 \beq{z1}
R_{12}(\bfu,z-w)R_{13}(\bfu-\hbar\gh_{(2)},z)R_{23}(\bfu,w)=
R_{23}(\bfu-\hbar\gh_{(1)},w)R_{13}(\bfu,z)R_{12}(\bfu-\hbar\gh_{(3)},z-w)\,.
 \eq


\subsubsection{Non-Dynamical Case: Baxter-Belavin-Drinfeld-Sklyanin vertex $R$-Matrix}

 The elliptic non-dynamical $R$-matrix is related to $\SLN$. It is defined  in the basis (\ref{B.11}) as follows:
 \beq{z2}
 R_{12}(z)=\sum\limits_{a\in\G_{N}}\vf_a(z,\om_a+\hbar)
T_a\otimes T_{-a}
 \eq
\begin{predl}
The Baxter-Belavin-Drinfeld-Sklyanin $R$-Matrix (\ref{z2}) satisfies the Quantum Yang-Baxter equation (\ref{z0})
\cite{Baxter,Belavin}.
\end{predl}

\underline{\emph{Proof}:} \vskip2mm

Consider a basic component of the tensor product $T_a\otimes T_{-a-b}\otimes T_b$:
 $$
\sum\limits_{c\in\G_{N}}\ka_{0,0}\ka_{a,b}\ka_{2a+2b,c}
\vf_{a-c}(\om_{a}-\om_{c}+\hbar,z-w)\vf_c(\om_{c}+\hbar,z)\vf_{-b-c}(-\om_{b}-\om_{c}+\hbar,w)=
 $$
 $$
\sum\limits_{c\in\G_{N}}\ka_{0,0}\ka_{b,a}\ka_{c,2a+2b}
\vf_{a-c}(\om_{a}-\om_{c}+\hbar,z-w)\vf_c(\om_{c}+\hbar,z)\vf_{-b-c}(-\om_{b}-\om_{c}+\hbar,w)
 $$
Making the shift $c\rightarrow -c-b$ in the l.h.s. we have the
following expression for (l.h.s.)-(r.h.s.):
 $$
\sum\limits_{c\in\G_{N}}\ka_{0,0}\ka_{b,a}\ka_{c,2a+2b}\Bigl(
\vf_{a+b+c}(\om_{a}+\om_{b}+\om_{c}+\hbar,z-w)
\vf_{-b-c}(-\om_{b}-\om_{c}+\hbar,z)\vf_{c}(\om_c+\hbar,w)\Bigr.-
 $$
 $$
\Bigl.\vf_{a-c}(\om_{a}-\om_{c}+\hbar,z-w)\vf_c(\om_c+\hbar,z)
\vf_{-b-c}(-\om_{b}-\om_{c}+\hbar,w)\Bigr)=
 $$
Using (\ref{ir}) in the case $a+b\neq 0\ \hbox{mod}\ \G_{N}$ we get:
 \beq{z22}
=\vf_a(\om_a+2\hbar,z)\vf_{-a-b}(-\om_{a}-\om_{b},w)\sum\limits_{c\in\G_{N}} \ka_{0,0}\ka_{b,a}\ka_{c,2a+2b}\times
\eq
 $$
\left(E_1(\om_{a}+\om_{b}+\om_{c}+\hbar)-E_1(\om_{a}-\om_{c}+\hbar)+
E_1(-\om_{b}-\om_{c}+\hbar)-E_1(\om_c+\hbar)\right)=0
 $$
Indeed, $\ka_{0,0}\ka_{b,a}\ka_{c,2a+2b}$ is invariant under the substitution $c\rightarrow c-a-b$.
 Then making this shift in the first $(E_1(\om_{a}+\om_{b}+\om_{c}+\hbar))$ and the third
$(E_1(-\om_{b}-\om_{c}+\hbar))$ terms one can see that the whole sum vanishes.

In the case $a+b=0\ \hbox{mod}\ \G_{N}$ it follows from (\ref{ir21}) for (l.h.s.)-(r.h.s.) that:
 $$
\vf_a(\om_a+2\hbar,z)\left(\frac{N}{2\pi l\sqrt{-1}}\right)^3 \sum\limits_{c\in\G_{N}}
\left(E_2(\om_c+\hbar)-E_2(\om_a-\om_c+\hbar)\right)=0\,,
 $$
where the normalization factor $\frac{N}{2\pi l\sqrt{-1}}=\ka(0,0)$ appears from (\ref{AA3a}). $\blacksquare$

\subsubsection{Dynamical Case: Felder $R$-Matrix}
The Felder $R$-matrix is defined as follows \cite{Fe1}:
  \beq{z3}
R_{12}(\bfu,z)=\sum\limits_{i,j}^{p} r_{ij}(\bfu,z)E_{ij}\otimes E_{ji} +\sum\limits_{\mu\neq\nu}^{N}
\rho_{\mu\nu}E_{\mu\mu}\otimes E_{\nu\nu}\,,
 \eq
  where
 $$
r_{ij}(\bfu,z) \equiv r_{ij}(z)=\phi(u_{ij}+\delta_{ij}\hbar,z),\ \ \ \rho_{ij}=\phi(-u_{ij},\hbar),\ \ \ u_{ij}=u_i-u_j
 $$
and
 $$
R_{13}(z,\bfu-\hbar\gh_{(2)})=\sum\limits_{m,n,s}^{N}
\check{r}_{mn}(z)E_{mn}\otimes \check{E}_{ss}\otimes E_{nm}
+\sum\limits_{\ga\neq\xi}^{N} \check{\rho}_{\ga\xi}E_{\ga\ga}\otimes
\check{E}_{ss}\otimes E_{\xi\xi}
 $$
We use "check" $\check{}$  to indicate the possible shift of the
argument of $R_{13}$ by $-\hbar\gh_{(2)}$:
 \beq{z4}
  \sum\limits_{m,n,s}^{N}\check{r}_{mn}(z)E_{mn}\otimes
\check{E}_{ss}\otimes E_{nm}=\sum\limits_{m,n,s}^{N}
\phi(u_{mn}+\delta_{mn}\hbar-\hbar\check{\delta}_{ms}+\hbar\check{\delta}_{ns},z)
E_{mn}\otimes \check{E}_{ss}\otimes E_{nm}
 \eq
 $$
\sum\limits_{\ga\neq\xi,s}^{N}
\check{\rho}_{\ga\xi}E_{\ga\ga}\otimes \check{E}_{ss}\otimes
E_{\xi\xi}=\sum\limits_{\ga\neq\xi,s}^{N}
\phi(-u_{\ga\xi}+\check{\delta}_{s\ga}\hbar-\check{\delta}_{s\xi}\hbar,\hbar)
E_{\ga\ga}\otimes \check{E}_{ss}\otimes E_{\xi\xi}
 $$
\begin{predl}
The Felder $R$-Matrix (\ref{z3}) satisfies the Quantum Dynamical Yang-Baxter equation (\ref{z1}).
\end{predl}

We omit here the proof of this Proposition since it is contained as a particular case of more general structure
 which will be
discussed in Section \ref{sec4}.



\subsubsection{Classical Limits}

Let us also make the classical limits of the quantum $R$-matrices (\ref{z2}),(\ref{z3}):
 \beq{x01}
r^{BD}_{12}(z,w)=\lim\limits_{\hbar\rightarrow 0}\left(R_{12}^{BD}(z,w)-\frac{1}{\hbar}1\otimes
1\right)\stackrel{(\ref{A.3a})}{=} E_1(z-w)1\otimes 1+\sum\limits_{\al\in\ti{\Gamma}_l}\vf_\al(z-w,\om_\al) T_\al\otimes
T_{-\al}
 \eq
Notice that the summation is taken over $\ti{\Gamma}_N$ (\ref{B.10}).
 This $r$-matrix satisfies the classical Yang-Baxter
equation:
 \beq{x011}
[r_{12},r_{13}]+[r_{12},r_{23}]+[r_{13},r_{23}]=0
 \eq
For the dynamical $r$-matrix we have:
 \beq{x02}
\begin{array}{c}
r^{F}_{12}(z,w)=\lim\limits_{\hbar\rightarrow
0}\left(R_{12}^{F}(z,w)-\frac{1}{\hbar}1\otimes
1\right)\stackrel{(\ref{A.3a})}{=}
\\
E_1(z-w)\sum\limits_{ii}E_{ii}\otimes E_{ii}+\sum\limits_{i\neq
j}^N\phi(z-w,u_{ij})E_{ij}\otimes E_{ji}-\sum\limits_{i\neq
j}^pE_1(u_{ij})\ E_{ii}\otimes E_{jj}
\end{array}
 \eq
The modified classical Yang-Baxter equation
 \beq{qyb}
[r_{12},r_{13}]+[r_{12},r_{23}]+[r_{13},r_{23}]+ D^{(1)}_\gh r_{23}-D^{(2)}_\gh r_{13}+D^{(3)}_\gh r_{12}=0,
 \eq
In the standard basis the operator $D^{(1)}_\gh$ is written as follows:
 $$
D^{(1)}_\gh=\sum\limits_{i=1}^N E_{ii}\otimes 1\otimes 1\ \p_{u_i}
 $$
It should be mentioned that the $r$-matrix (\ref{x02}) without the last on term also satisfies (\ref{qyb}). The reason is
that it can be removed by the dynamical twist (see e.g. \cite{ell_qg5} or \cite{LOSZ3}).

Note, that the both $r$-matrices $r^{BD}$ (\ref{x01}) and $ r^{F}$ (\ref{x02}) are particular cases of the general form
(\ref{Rma}).



\subsection{General  Quantum $R$-Matrices for $\SLN$-Bundles}\label{sec4}

Let us consider the basis (\ref{o04}) in $\GLN\simeq {{\rm GL}(p,{\mathbb C})}\times{{\rm GL}(l,{\mathbb C})}$,
where $N=lp,\
l,p\in\mZ$:
 \beq{x1}
  E^a_{ij}=E_{ij}\otimes T_a\,,\ \ E_{ij}\in{{\rm gl}(p,{\mathbb C})}\,,
  \ \ T_a\in {{\rm gl}(l,{\mathbb
C})}\,,
 \eq
where $E_{ij}$ is a standard matrix basis in the fundamental representation of ${{\rm GL}(p,{\mathbb C})}$ and
$T_a$ is the
basis of ${{\rm GL}(l,{\mathbb C})}$ defined in (\ref{B.11}). From (\ref{AA3a}) it follows that:
 \beq{x2}
 E^a_{ij}E^b_{kl}=\ka_{a,b}\delta_{kj}E^{a+b}_{il}
 \eq
Now let us introduce the following $R$-matrix:
 \beq{x31}
 R_{12}(\bfu,z)=\sum\limits_{i,j}^{p}\sum\limits_{a\in
\Gamma_l} r_{ij}^a(\bfu,z)E_{ij}^a\otimes E_{ji}^{-a} +\sum\limits_{\mu\neq\nu}^{p} \rho_{\mu\nu}^0 E_{\mu\mu}^0\otimes
E_{\nu\nu}^0\footnote{We will also use notation $\delta_{a,0}\rho_{ij}^a=\rho_{ij}^0$ in order to keep uniformity in
formulae.}\,,
 \eq
where
 $$
r_{ij}^a(\bfu,z) \equiv r_{ij}^a(z)=\vf_{-a}(-u_{ij}-\delta_{ij}\hbar,z)\,,~
\rho_{ij}^0=\phi(-lu_{ij},l\hbar)\,,
$$
$$
u_{ij}=u_i-u_j\,, ~\om_a=\frac{a_1+\tau a_2}l\,.
 $$

In particular, if $p=1$, $(l=N)$ then (\ref{x31}) coincides with (\ref{z2} ). For $p=N$ $(l=1)$ we come to (\ref{z3}). In
this way the elliptic  $R$-matrix   (\ref{x31}) unifies the dynamical and non-dynamical cases.

Consider dependence of $R(\bfu,z)$ on $z$ and $\bfu$.
The shifts of $z$ yield (see (\ref{92}))
$$
R(\bfu,z+1)=Q_{p\times p}R(\bfu,z)Q^{-1}_{p\times p}\,,~~~
R(\bfu,z+\tau)=\bfe_l(\bfu+\hbar)\La_{p\times p}R(\bfu,z)\La_{p\times p}^{-1}\bfe_l(-\bfu-\hbar)\,,
$$
where $Q_{p\times p}$ and $\La_{p\times p}$ act on the second factor of the basis (\ref{x1}), and
the adjoint transformation by
$\bfe_l(\bfu+\hbar)$ acts on the first factor as  $E_{ij}\to\bfe_l(u_i-u_j+\de_{ij}\hbar)E_{ij}$.
This conditions define the characteristic class of bundle.

Consider  quasi-periodicities of $R(\bfu,z)$, $(\bfu\in\ti\gh_0)$
 with respect shifts of the weight lattice  $P_l$ or the root lattice $Q_l$
in $\ti\gh_0$ (see (\ref{cl1}), (\ref{clp}). Let $\ga=(m_1,\ldots,m_p)\in P_l,$ or $Q_l$ and
$\Upsilon_p(\ga,z)=\bfe(\ga z)=\di(\bfe(m_1z),\ldots,\bfe(m_pz))$
It follows from (\ref{x31}) and (\ref{qpu}) that
$$
R(\bfu+\ga,z)=R(\bfu,z)\,,~~R(\bfu+\tau\ga,z)=(\Upsilon^{-1}_p(\ga,z)\otimes Id_l)R(\bfu,z)(\Upsilon_p(\ga,z)\otimes Id_l)\,.
$$
It means that $R(\bfu,z)$ is a section of trivial bundle over the moduli spaces (\ref{cl1}), (\ref{clp}).

\bigskip

Now we prove the main result of the paper:
\begin{theor}\label{theor}
The $R$-Matrix (\ref{x31}) satisfies the Quantum Dynamical Yang-Baxter equation (\ref{z1}).
\end{theor}

\underline{\emph{Proof}:} \vskip2mm

By analogy with the notation "check" $\check{}$ from (\ref{z4}) we
use "acute" $\acute{}$ for the indication of possible shift of the
argument of $\acute{R}_{23}$ (in the r.h.s. of (\ref{z1})) and
"tilde" $\tilde{}$ for $\tilde{R}_{12}$ (in the r.h.s. of
(\ref{z1})). Let us write down equation (\ref{z1}) explicitly:
 \beq{z5}
\begin{array}{c}
l.h.s.=\sum \\
r_{ij}^{a-c}(z-w)\hhr_{mn}^{c}(z)r_{kl}^{-b-c}(w)\
E_{ij}^{a-c}E_{mn}^{c}\otimes
E_{ji}^{c-a}\hhe_{ss}^{0}E_{kl}^{-b-c}\otimes
E_{mn}^{-c}E_{kl}^{b+c}+
\\
r_{ij}^{a-c}(z-w)\hhr_{mn}^{c}(z)\rho_{\al\be}^{0}(w)\
E_{ij}^{a-c}E_{mn}^{c}\otimes
E_{ji}^{c-a}\hhe_{ss}^{0}E_{\al\al}^{0}\otimes
E_{mn}^{-c}E_{\be\be}^{0}+
\\
r_{ij}^{a-c}(z-w)\check{\rho}_{\ga\xi}^{0}(z)r_{kl}^{-b-c}(w)\
E_{ij}^{a-c}E_{\ga\ga}^{0}\otimes
E_{ji}^{c-a}\hhe_{ss}^{0}E_{kl}^{-b-c}\otimes
E_{\xi\xi}^{0}E_{kl}^{b+c}+
\\
\rho_{\mu\nu}^{0}(z-w)\hhr_{mn}^{c}(z)r_{kl}^{-b-c}(w)\
E_{\mu\mu}^{0}E_{mn}^{c}\otimes
E_{\nu\nu}^{0}\hhe_{ss}^{0}E_{kl}^{-b-c}\otimes
E_{mn}^{-c}E_{kl}^{b+c}+
\\
\
\\
r_{ij}^{a-c}(z-w)\check{\rho}_{\ga\xi}^{0}(z)\rho_{\al\be}^{0}(w)\
E_{ij}^{a-c}E_{\ga\ga}^{0}\otimes
E_{ji}^{c-a}\hhe_{ss}^{0}E_{\al\al}^{0}\otimes
E_{\xi\xi}^{0}E_{\be\be}^{0}+
\\
\rho_{\mu\nu}^{0}(z-w)\hhr_{mn}^{c}(z)\rho_{\al\be}^{0}(w)\
E_{\mu\mu}^{0}E_{mn}^{c}\otimes
E_{\nu\nu}^{0}\hhe_{ss}^{0}E_{\al\al}^{0}\otimes
E_{mn}^{-c}E_{\be\be}^{0}+
\\
\rho_{\mu\nu}^{0}(z-w)\check{\rho}_{\ga\xi}^{0}(z)r_{kl}^{-b-c}(w)\
E_{\mu\mu}^{0}E_{\ga\ga}^{0}\otimes
E_{\nu\nu}^{0}\hhe_{ss}^{0}E_{kl}^{-b-c}\otimes
E_{\xi\xi}^{0}E_{kl}^{b+c}+
\\
\rho_{\mu\nu}^{0}(z-w)\check{\rho}_{\ga\xi}^{0}(z)\rho_{\al\be}^{0}(w)\
E_{\mu\mu}^{0}E_{\ga\ga}^{0}\otimes
E_{\nu\nu}^{0}\hhe_{ss}^{0}E_{\al\al}^{0}\otimes
E_{\xi\xi}^{0}E_{\be\be}^{0}=
\end{array}
 \eq
 \beq{z6}
\begin{array}{c}
r.h.s.=\sum
\\
\ttr_{ij}^{a-c}(z-w)r_{mn}^{c}(z) \aar_{kl}^{-b-c}(w)\
\aae_{qq}^{0}E_{mn}^{c}E_{ij}^{a-c}\otimes
E_{kl}^{-b-c}E_{ji}^{c-a}\otimes
E_{lk}^{b+c}E_{nm}^{-c}\tte_{tt}^{0}+
\\
\ttr_{ij}^{a-c}(z-w)r_{mn}^{c}(z)\acute{\rho}_{\al\be}^{0}(w)\
\aae_{qq}^{0}E_{mn}^{c}E_{ij}^{a-c}\otimes
E_{\al\al}^{0}E_{ji}^{c-a}\otimes
E_{\be\be}^{0}E_{nm}^{-c}\tte_{tt}^{0}+
\\
\ttr_{ij}^{a-c}(z-w)\rho_{\ga\xi}^{0}(z) \aar_{kl}^{-b-c}(w)\
\aae_{qq}^{0}E_{\ga\ga}^{0}E_{ij}^{a-c}\otimes
E_{kl}^{-b-c}E_{ji}^{c-a}\otimes
E_{lk}^{b+c}E_{\xi\xi}^{0}\tte_{tt}^{0}+
\\
\tilde{\rho}_{\mu\nu}^{0}(z-w)r_{mn}^{c}(z) \aar_{kl}^{-b-c}(w)\
\aae_{qq}^{0}E_{mn}^{c}E_{\mu\mu}^{0}\otimes
E_{kl}^{-b-c}E_{\nu\nu}^{0}\otimes
E_{lk}^{b+c}E_{nm}^{-c}\tte_{tt}^{0}+
\\
\
\\
\ttr_{ij}^{a-c}(z-w)\rho_{\ga\xi}^{0}(z)
\acute{\rho}_{\al\be}^{0}(w)\
\aae_{qq}^{0}E_{\ga\ga}^{0}E_{ij}^{a-c}\otimes
E_{\al\al}^{0}E_{ji}^{c-a}\otimes
E_{\be\be}^{0}E_{\xi\xi}^{0}\tte_{tt}^{0}+
\\
\tilde{\rho}_{\mu\nu}^{0}(z-w)r_{mn}^{c}(z)
\acute{\rho}_{\al\be}^{0}(w)\
\aae_{qq}^{0}E_{mn}^{c}E_{\mu\mu}^{0}\otimes
E_{\al\al}^{0}E_{\nu\nu}^{0}\otimes
E_{\be\be}^{0}E_{nm}^{-c}\tte_{tt}^{0}+
\\
\tilde{\rho}_{\mu\nu}^{0}(z-w)\rho_{\ga\xi}^{0}(z)
\aar_{kl}^{-b-c}(w)\
\aae_{qq}^{0}E_{\ga\ga}^{0}E_{\mu\mu}^{0}\otimes
E_{kl}^{-b-c}E_{\nu\nu}^{0}\otimes
E_{lk}^{b+c}E_{\xi\xi}^{0}\tte_{tt}^{0}+
\\
\tilde{\rho}_{\mu\nu}^{0}(z-w)\rho_{\ga\xi}^{0}(z)
\acute{\rho}_{\al\be}^{0}(w)\
\aae_{qq}^{0}E_{\ga\ga}^{0}E_{\mu\mu}^{0}\otimes
E_{\al\al}^{0}E_{\nu\nu}^{0}\otimes
E_{\be\be}^{0}E_{\xi\xi}^{0}\tte_{tt}^{0},
\end{array}
 \eq where the sums are taken over all indices\footnote{Here and
elsewhere we shall omit the indices and limits of summation when it can be done without ambiguity }.
Using (\ref{x2}) and
(\ref{kap}) we get
 \beq{z7}
\begin{array}{c}
l.h.s.= \sum
\\
\ka_{0,0}\ka_{a,b}\ka_{2a+2b,c}\ \check{\delta}_{is}\
r_{ij}^{a-c}(z-w)\hhr_{jn}^{c}(z)r_{ij}^{-b-c}(w)\ E^{a}_{in}\otimes
E_{jj}^{-a-b}\otimes E_{ni}^{b}+
\\
\delta_{b,-c}\ \ka_{0,0}\ka_{a,b}\ka_{2a+2b,c}\ \check{\delta}_{is}\
r_{ij}^{a-c}(z-w)\hhr_{jn}^{c}(z)\rho_{ij}^{-b-c}(w)\
E^{a}_{in}\otimes E_{ji}^{-a-b}\otimes E_{nj}^{b}+
\\
\delta_{0,c}\ \ka_{0,0}\ka_{a,b}\ka_{2a+2b,c}\ \check{\delta}_{is}\
r_{ij}^{a-c}(z-w)\check{\rho}_{jl}^{c}(z)r_{jl}^{-b-c}(w)\
E^{a}_{ij}\otimes E_{jl}^{-a-b}\otimes E_{li}^{b}+
\\
\delta_{a,c}\ \ka_{0,0}\ka_{a,b}\ka_{2a+2b,c}\ \check{\delta}_{ks}\
\rho_{mk}^{a-c}(z-w)\hhr_{mn}^{c}(z)r_{km}^{-b-c}(w)\
E^{a}_{mn}\otimes E_{km}^{-a-b}\otimes E_{nk}^{b}+
\\
\
\\
\delta_{0,c}\delta_{b,-c}\ \ka_{0,0}\ka_{a,b}\ka_{2a+2b,c}\
\check{\delta}_{is}\
r_{ij}^{a-c}(z-w)\check{\rho}_{j\xi}^{c}(z)\rho_{i\xi}^{-b-c}(w)\
E^{a}_{ij}\otimes E_{ji}^{-a-b}\otimes E_{\xi\xi}^{b}+
\\
\delta_{a,c}\delta_{b,-c}\ \ka_{0,0}\ka_{a,b}\ka_{2a+2b,c}\
\check{\delta}_{\al s}\
\rho_{m\al}^{a-c}(z-w)\hhr_{mn}^{c}(z)\rho_{\al m}^{-b-c}(w)\
E^{a}_{mn}\otimes E_{\al\al}^{-a-b}\otimes E_{nm}^{b}+
\\
\delta_{0,c}\delta_{a,c}\ \ka_{0,0}\ka_{a,b}\ka_{2a+2b,c}\
\check{\delta}_{ks}\ \rho_{\ga k}^{a-c}(z-w)\check{\rho}_{\ga
l}^{c}(z)r_{kl}^{-b-c}(w)\ E^{a}_{\ga\ga}\otimes
E_{kl}^{-a-b}\otimes E_{lk}^{b}+
\\
\delta_{0,c}\delta_{b,-c}\delta_{a,c}\
\ka_{0,0}\ka_{a,b}\ka_{2a+2b,c}\ \check{\delta}_{\al s}\
\rho_{\mu\al}^{a-c}(z-w)\check{\rho}_{\mu\be}^{c}(z)\rho_{\al\be}^{-b-c}(w)\
E^{a}_{\mu\mu}\otimes E_{\al\al}^{-a-b}\otimes E_{\be\be}^{b}=
\end{array}
 \eq
 \beq{z8}
\begin{array}{c}
r.h.s.=  \sum
\\
\ka_{0,0}\ka_{b,a}\ka_{c,2a+2b}\
\acute{\delta}_{qm}\tilde{\delta}_{tm}\
\ttr_{kj}^{a-c}(z-w)r_{mk}^{c}(z)\aar_{kj}^{-b-c}(w)\
E^{a}_{mj}\otimes E_{kk}^{-a-b}\otimes E_{jm}^{b}+
\\
\delta_{b,-c}\ \ka_{0,0}\ka_{b,a}\ka_{c,2a+2b}\
\acute{\delta}_{qm}\tilde{\delta}_{tm}\
\ttr_{ij}^{a-c}(z-w)r_{mi}^{c}(z)\acute{\rho}_{ji}^{-b-c}(w)\
E^{a}_{mj}\otimes E_{ji}^{-a-b}\otimes E_{im}^{b}+
\\
\delta_{0,c}\ \ka_{0,0}\ka_{b,a}\ka_{c,2a+2b}\
\acute{\delta}_{qi}\tilde{\delta}_{tk}\
\ttr_{ij}^{a-c}(z-w)\rho_{ik}^{c}(z)\aar_{kj}^{-b-c}(w)\
E^{a}_{ij}\otimes E_{ki}^{-a-b}\otimes E_{jk}^{b}+
\\
\delta_{a,c}\ \ka_{0,0}\ka_{b,a}\ka_{c,2a+2b}\
\acute{\delta}_{qm}\tilde{\delta}_{tm}\
\tilde{\rho}_{nl}^{a-c}(z-w)r_{mn}^{c}(z)\aar_{nl}^{-b-c}(w)\
E^{a}_{mn}\otimes E_{nl}^{-a-b}\otimes E_{lm}^{b}+
\\
\
\\
\delta_{0,c}\delta_{b,-c}\ \ka_{0,0}\ka_{b,a}\ka_{c,2a+2b}\
\acute{\delta}_{qi}\tilde{\delta}_{t\be}\
\ttr_{ij}^{a-c}(z-w)\rho_{i\be}^{c}(z)\acute{\rho}_{j\be}^{-b-c}(w)\
E^{a}_{ij}\otimes E_{ji}^{-a-b}\otimes E_{\be\be}^{b}+
\\
\delta_{b,-c}\delta_{a,c}\ \ka_{0,0}\ka_{b,a}\ka_{c,2a+2b}\
\acute{\delta}_{qm}\tilde{\delta}_{tm}\
\tilde{\rho}_{n\al}^{a-c}(z-w)r_{mn}^{c}(z)\acute{\rho}_{\al
n}^{-b-c}(w)\ E^{a}_{mn}\otimes E_{\al\al}^{-a-b}\otimes E_{nm}^{b}+
\\
\delta_{0,c}\delta_{a,c}\ \ka_{0,0}\ka_{b,a}\ka_{c,2a+2b}\
\acute{\delta}_{q\ga}\tilde{\delta}_{tk}\ \tilde{\rho}_{\ga
l}^{a-c}(z-w)\rho_{\ga k}^{c}(z)\aar_{kl}^{-b-c}(w)\
E^{a}_{\ga\ga}\otimes E_{kl}^{-a-b}\otimes E_{lk}^{b}+
\\
\delta_{0,c}\delta_{b,-c}\delta_{a,c}\
\ka_{0,0}\ka_{b,a}\ka_{c,2a+2b}\
\acute{\delta}_{q\ga}\tilde{\delta}_{t\be}\
\tilde{\rho}_{\ga\al}^{a-c}(z-w)\rho_{\ga\be}^{c}(z)\acute{\rho}_{\al\be}^{-b-c}(w)\
E^{a}_{\ga\ga}\otimes E_{\al\al}^{-a-b}\otimes E_{\be\be}^{b}
\end{array}
 \eq
The notation $\tilde{\delta}_{tm}$ here (for example, in the first
line of (\ref{z8})) means that the corresponding $\ttr_{kj}$ has the
shift of the argument $u_{kj}$ by $-\hbar$ if $k=m$ and by $+\hbar$
when $j=m$.

Making change of summation variable $c\rightarrow -c-b$ in the
l.h.s. (\ref{z7}) we get the same factor
$\ka_{0,0}\ka_{a,b}\ka_{2a+2b,c}\rightarrow\ka_{0,0}\ka_{b,a}\ka_{c,2a+2b}$
for both sides due to (\ref{kap}):
 \beq{z9}
\begin{array}{c}
l.h.s.= \sum
\\
\ka_{0,0}\ka_{b,a}\ka_{c,2a+2b}\ \check{\delta}_{is}\
r_{ij}^{a+b+c}(z-w)\hhr_{jn}^{-b-c}(z)r_{ij}^{c}(w)\
E^{a}_{in}\otimes E_{jj}^{-a-b}\otimes E_{ni}^{b}+
\\
\delta_{0,c}\ \ka_{0,0}\ka_{b,a}\ka_{c,2a+2b}\ \check{\delta}_{is}\
r_{ij}^{a+b+c}(z-w)\hhr_{jn}^{-b-c}(z)\rho_{ij}^{c}(w)\
E^{a}_{in}\otimes E_{ji}^{-a-b}\otimes E_{nj}^{b}+
\\
\delta_{b,-c}\ \ka_{0,0}\ka_{b,a}\ka_{c,2a+2b}\ \check{\delta}_{is}\
r_{ij}^{a+b+c}(z-w)\check{\rho}_{jl}^{-b-c}(z)r_{jl}^{c}(w)\
E^{a}_{ij}\otimes E_{jl}^{-a-b}\otimes E_{li}^{b}+
\\
\delta_{a+b,-c}\ \ka_{0,0}\ka_{b,a}\ka_{c,2a+2b}\
\check{\delta}_{ks}\
\rho_{mk}^{a+b+c}(z-w)\hhr_{mn}^{-b-c}(z)r_{km}^{c}(w)\
E^{a}_{mn}\otimes E_{km}^{-a-b}\otimes E_{nk}^{b}+
\\
\
\\
\delta_{0,c}\delta_{b,-c}\ \ka_{0,0}\ka_{b,a}\ka_{c,2a+2b}\
\check{\delta}_{is}\
r_{ij}^{a+b+c}(z-w)\check{\rho}_{j\xi}^{-b-c}(z)\rho_{i\xi}^{c}(w)\
E^{a}_{ij}\otimes E_{ji}^{-a-b}\otimes E_{\xi\xi}^{b}+
\\
\delta_{a+b,-c}\delta_{0,c}\ \ka_{0,0}\ka_{b,a}\ka_{c,2a+2b}\
\check{\delta}_{\al s}\
\rho_{m\al}^{a+b+c}(z-w)\hhr_{mn}^{-b-c}(z)\rho_{\al m}^{c}(w)\
E^{a}_{mn}\otimes E_{\al\al}^{-a-b}\otimes E_{nm}^{b}+
\\
\delta_{b,-c}\delta_{a+b,-c}\ \ka_{0,0}\ka_{b,a}\ka_{c,2a+2b}\
\check{\delta}_{ks}\ \rho_{\ga k}^{a+b+c}(z-w)\check{\rho}_{\ga
l}^{-b-c}(z)r_{kl}^{c}(w)\ E^{a}_{\ga\ga}\otimes
E_{kl}^{-a-b}\otimes E_{lk}^{b}+
\\
\delta_{0,c}\delta_{b,-c}\delta_{a+b,-c} \ka_{0,0}\ka_{b,a}\ka_{c,2a+2b}\ \check{\delta}_{\al s}
\rho_{\mu\al}^{a+b+c}(z-w)\check{\rho}_{\mu\be}^{-b-c}(z)\rho_{\al\be}^{c}(w)\ E^{a}_{\mu\mu}\otimes E_{\al\al}^{-a-b}\otimes
E_{\be\be}^{b}
\end{array}
 \eq
 \beq{z10}
\begin{array}{c}
r.h.s.=  \sum
\\
\ka_{0,0}\ka_{b,a}\ka_{c,2a+2b}\
\acute{\delta}_{qm}\tilde{\delta}_{tm}\
\ttr_{kj}^{a-c}(z-w)r_{mk}^{c}(z)\aar_{kj}^{-b-c}(w)\
E^{a}_{mj}\otimes E_{kk}^{-a-b}\otimes E_{jm}^{b}+
\\
\delta_{b,-c}\ \ka_{0,0}\ka_{b,a}\ka_{c,2a+2b}\
\acute{\delta}_{qm}\tilde{\delta}_{tm}\
\ttr_{ij}^{a-c}(z-w)r_{mi}^{c}(z)\acute{\rho}_{ji}^{-b-c}(w)\
E^{a}_{mj}\otimes E_{ji}^{-a-b}\otimes E_{im}^{b}+
\\
\delta_{0,c}\ \ka_{0,0}\ka_{b,a}\ka_{c,2a+2b}\
\acute{\delta}_{qi}\tilde{\delta}_{tk}\
\ttr_{ij}^{a-c}(z-w)\rho_{ik}^{c}(z)\aar_{kj}^{-b-c}(w)\
E^{a}_{ij}\otimes E_{ki}^{-a-b}\otimes E_{jk}^{b}+
\\
\delta_{a,c}\ \ka_{0,0}\ka_{b,a}\ka_{c,2a+2b}\
\acute{\delta}_{qm}\tilde{\delta}_{tm}\
\tilde{\rho}_{nl}^{a-c}(z-w)r_{mn}^{c}(z)\aar_{nl}^{-b-c}(w)\
E^{a}_{mn}\otimes E_{nl}^{-a-b}\otimes E_{lm}^{b}+
\\
\
\\
\delta_{0,c}\delta_{b,-c}\ \ka_{0,0}\ka_{b,a}\ka_{c,2a+2b}\
\acute{\delta}_{qi}\tilde{\delta}_{t\be}\
\ttr_{ij}^{a-c}(z-w)\rho_{i\be}^{c}(z)\acute{\rho}_{j\be}^{-b-c}(w)\
E^{a}_{ij}\otimes E_{ji}^{-a-b}\otimes E_{\be\be}^{b}+
\\
\delta_{b,-c}\delta_{a,c}\ \ka_{0,0}\ka_{b,a}\ka_{c,2a+2b}\
\acute{\delta}_{qm}\tilde{\delta}_{tm}\
\tilde{\rho}_{n\al}^{a-c}(z-w)r_{mn}^{c}(z)\acute{\rho}_{\al
n}^{-b-c}(w)\ E^{a}_{mn}\otimes E_{\al\al}^{-a-b}\otimes E_{nm}^{b}+
\\
\delta_{0,c}\delta_{a,c}\ \ka_{0,0}\ka_{b,a}\ka_{c,2a+2b}\
\acute{\delta}_{q\ga}\tilde{\delta}_{tk}\ \tilde{\rho}_{\ga
l}^{a-c}(z-w)\rho_{\ga k}^{c}(z)\aar_{kl}^{-b-c}(w)\
E^{a}_{\ga\ga}\otimes E_{kl}^{-a-b}\otimes E_{lk}^{b}+
\\
\delta_{0,c}\delta_{b,-c}\delta_{a,c}\
\ka_{0,0}\ka_{b,a}\ka_{c,2a+2b}\
\acute{\delta}_{q\ga}\tilde{\delta}_{t\be}\
\tilde{\rho}_{\ga\al}^{a-c}(z-w)\rho_{\ga\be}^{c}(z)\acute{\rho}_{\al\be}^{-b-c}(w)\
E^{a}_{\ga\ga}\otimes E_{\al\al}^{-a-b}\otimes E_{\be\be}^{b}
\end{array}
 \eq
A careful check shows that the equality (\ref{z9}-\ref{z10}) holds. The general idea of the verification is the following: if
$a\neq -b$ the proof is similar to the one given for the non-dynamical case (\ref{z2}) and if $a=-b$ the equality is achieved
by "$\rho$"-term in the $R$-matrix via summation of (\ref{a95}) over $c$.

Let us demonstrate the verification for some concrete cases:

\vspace{2mm}

$\underline{E_{ij}^a\otimes E_{kk}^{-a-b}\otimes E_{ji}^b:
}$\footnote{Here and elsewhere we imply unequal lower indices while
the upper may be dependent, e.g.  $a=-b$ or $a=0$ or $b=0$. Note
also that the summution will be taken only over $c$.}

 \beq{z11}
\begin{array}{c}
l.h.s.=\sum\limits_{c\in\Gamma_l}\ka_{0,0}\ka_{b,a}\ka_{c,2a+2b}\left(\right.
\\
\left.\check{\delta}_{is}r_{ik}^{a+b+c}(z-w)\check{r}_{kj}^{-b-c}(z)r_{ik}^c(w)+
\check{\delta}_{ks}\delta_{a+b,-c}\delta_{0,c}\
\rho_{ik}^{a+b+c}(z-w)\check{r}_{ij}^{-b-c}(z)\rho_{ki}^c(w)
\right)=
\\
r.h.s.=\sum\limits_{c\in\Gamma_l}\ka_{0,0}\ka_{b,a}\ka_{c,2a+2b}\left(\right.
\\
\left.\tilde{\delta}_{ti}\acute{\delta}_{qi}
\tilde{r}_{kj}^{a-c}(z-w){r}_{ik}^{c}(z)\acute{r}_{kj}^{-b-c}(w)+
\tilde{\delta}_{ti}\acute{\delta}_{qi}\delta_{a,c}\delta_{b,-c}\
\tilde{\rho}_{jk}^{a-c}(z-w){r}_{ij}^{c}(z)\acute{\rho}_{kj}^{-b-c}(w)
\right)
\end{array}
 \eq

Index $s$ in the l.h.s. of (\ref{z11}) is responsible for the possible shift of argument in $\check{r}$.
 In this case we can
see that it does not match the corresponding arguments. The same holds for indices $t$ and $q$.
Thus there are no any shifts
in this case. Now combining the first terms from both sides we get:
 \beq{z12}
\begin{array}{c}
\sum\limits_{c\in\Gamma_l}\ka_{0,0}\ka_{b,a}\ka_{c,2a+2b}\left(\right.
\\
\left.\vf_{a+b+c}(z-w,u_{ik}+\om_a+\om_b+\om_c)\vf_{-b-c}(z,u_{kj}-\om_b-\om_c)
\vf_c(w,u_{ik}+\om_c)\right.-
\\
\left.\vf_{a-c}(z-w,u_{kj}+\om_a-\om_c)\vf_{c}(z,u_{ik}+\om_c)
\vf_{-b-c}(w,u_{kj}-\om_b-\om_c)\right)=\delta_{a,-b}{\ka_{0,0}^3}{l^2}
\left(\right.
\\
\
\\
\left.\phi(l\hbar,-lu_{jk})\vf_a(z,u_{ij}+\om_a)\phi(l\hbar,-lu_{kj})-
\phi(l\hbar,-lu_{ik})\vf_a(z,u_{ij}+\om_a)\phi(l\hbar,-lu_{ki})
\right)
\end{array}
 \eq

Let us examine the l.h.s. of (\ref{z12}). Due to (\ref{a95}) it
simplifies to:

1. For $a\neq -b\ \hbox{mod}\ \mZ^{(2)}_N$:
 $$
\begin{array}{c}
l.h.s. (\ref{z12})=\vf_a(z,\om_a+u_{ij})\vf_{-a-b}(w,-\om_a-\om_b)
\sum\limits_{c\in\Gamma_l}\ka_{0,0}\ka_{b,a}\ka_{c,2a+2b}
\left(\right.
\\
\left.E_1(u_{ik}+\om_a+\om_b+\om_c)-E_1(u_{kj}+\om_a-\om_c)+
E_1(u_{kj}-\om_b-\om_c)-E_1(u_{ik}+\om_c)\right)=0
\end{array}
 $$
exactly as in (\ref{z22}).

2. For $a=-b\ \hbox{mod}\ \mZ^{(2)}_N$:  \beq{z13}
\begin{array}{c}
\delta_{a,-b}{\ka_{0,0}^3}\vf_a(z,u_{ij}+\om_a)\sum\limits_{c\in\Gamma_l}
(E_2(u_{ik}+\om_c)-E_2(u_{kj}+\om_a-\om_c))\stackrel{(\ref{a93})}{=}
\\
\delta_{a,-b}{\ka_{0,0}^3}\vf_a(z,u_{ij}+\om_a)l^2
(E_2(lu_{ik})-E_2(lu_{kj}))\stackrel{(\ref{ad21})}{=}r.h.s.
(\ref{z12})
\end{array}
 \eq

$\underline{E_{ij}^a\otimes E_{ii}^{-a-b}\otimes E_{ji}^b: }$

 \beq{z14}
\begin{array}{c}
l.h.s.=\sum\limits_{c\in\Gamma_l}\ka_{0,0}\ka_{b,a}\ka_{c,2a+2b}
\check{\delta}_{is}r_{ii}^{a+b+c}(z-w)\check{r}_{ij}^{-b-c}(z)r_{ii}^c(w)=
\\
r.h.s.=\sum\limits_{c\in\Gamma_l}\ka_{0,0}\ka_{b,a}\ka_{c,2a+2b}\left(\right.
\\
\left.\tilde{\delta}_{ti}\acute{\delta}_{qi}
\tilde{r}_{ij}^{a-c}(z-w){r}_{ii}^{c}(z)\acute{r}_{ij}^{-b-c}(w)+
\delta_{b,-c}\delta_{a,c}\ \acute{\delta}_{qi}\tilde{\delta}_{ti}\
\tilde{\rho}_{ji}^{a-c}(z-w)r_{ij}^{c}(z)\acute{\rho}_{ij}^{-b-c}(w)
\right)
\end{array}
 \eq

Notice that the shift of $u$ takes place in all arguments. Further check is similar to the previous case. Indeed:
 \beq{z15}
\begin{array}{c}
\sum\limits_{c\in\Gamma_l}\ka_{0,0}\ka_{b,a}\ka_{c,2a+2b}\left(\right.
\\
\left.\vf_{a+b+c}(z-w,\om_a+\om_b+\om_c+\hbar)\vf_{-b-c}(z,u_{ij}-\om_b-\om_c-\hbar)
\vf_c(w,\om_c+\hbar)\right.-
\\
\left.\vf_{a-c}(z-w,u_{ij}+\om_a-\om_c-\hbar)\vf_{c}(z,\om_c+\hbar)
\vf_{-b-c}(w,u_{ij}-\om_b-\om_c-\hbar)\right)=
\\
\delta_{a,-b}{\ka_{0,0}^3}{l^2}
\phi(l\hbar,-lu_{ji}-l\hbar)\vf_a(z,u_{ij}+\om_a)\phi(l\hbar,-lu_{ij}+l\hbar)
\end{array}
 \eq

$\underline{E_{ij}^a\otimes E_{ki}^{-a-b}\otimes E_{jk}^b: }$

 \beq{z16}
\begin{array}{c}
\delta_{c,0}\vf_{a+b+c}(z-w,u_{ik}+\om_a+\om_b+\om_c)\vf_{-b-c}(z,u_{kj}-\om_b-\om_c)
\phi(l\hbar,-lu_{ik})+
\\
\delta_{c,-a-b}\phi(l\hbar,-lu_{ik})
\vf_{-b-c}(z,u_{ij}-\om_b-\om_c) \vf_{c}(w,u_{ki}+\om_c)=
\\
\delta_{c,0}\vf_{a-c}(z-w,u_{ij}+\om_a-\om_c)\phi(l\hbar,-lu_{ik})\vf_{-b-c}(w,u_{kj}-\om_b-\om_c)
\end{array}
 \eq
or
 \beq{z17}
\begin{array}{c}
\vf_{a+b}(z-w,u_{ik}+\om_a+\om_b)\vf_{-b}(z,u_{kj}-\om_b) +
\\
\vf_{a}(z,u_{ij}+\om_a)
\vf_{-a-b}(w,u_{ki}-\om_a-\om_b)\stackrel{(\ref{a94})}{=}
\\
\vf_{a}(z-w,u_{ij}+\om_a)\vf_{-b}(w,u_{kj}-\om_b)
\\
\ \ \ \ \ \ \ \ \ \ \ \ \ \ \ \ \ \ \ \ \ \ \ \ \ \ \ \ \ \ \ \ \ \ \ \ \ \ \ \ \ \ \ \ \ \ \ \ \ \ \ \ \ \ \ \ \ \ \ \ \ \ \
\ \ \ \ \ \ \ \ \ \ \ \ \ \ \ \ \ \ \ \ \blacksquare
\end{array}
 \eq


\subsection{Trigonometric and Rational Limits}
We can calculate the trigonometric limit $ \Im m\tau\to+\infty$ of the elliptic $R$-matrix (\ref{x31})  using (\ref{A.3z})
and (\ref{A.3zz})
$$
R^{trig}(\bfu,z)=\sum\limits_{i,j}^{p}\sum\limits_{a\in
\Gamma_l} r_{ij}^a(\bfu,z)E_{ij}^a\otimes E_{ji}^{-a} +\sum\limits_{\mu\neq\nu}^{p} \rho_{\mu\nu}^0 E_{\mu\mu}^0\otimes
E_{\nu\nu}^0\footnote{We will also use notation $\delta_{a,0}\rho_{ij}^a=\rho_{ij}^0$ in order to keep uniformity in
formulae.}\,,
 $$
 $$
r_{ij}^a(\bfu,z)
=\left\{
 \begin{array}{cc}
\cot\pi z+\cot\pi(u_{ij}+\frac{a_1}N+\delta_{ij}\hbar)    & a_2=0\,, \\
   \bfe((\frac{a_2}N+1)z)\sin^{-1}\pi z & a_2\neq 0\,.
 \end{array}
 \right.
$$
$$
\rho_{ij}^0=\frac{\sin\pi (l\hbar-lu_{ij})}{\sin\pi (l\hbar)\sin\pi (lu_{ji})}\,, \ \
u_{ij}=u_i-u_j\,.
 $$

Going to the rational limit we find
$$
r_{ij}^a(\bfu,z) =
\left\{
 \begin{array}{cc}
\f1{\pi z}+\f1{\pi(u_{ij}+\frac{a_1}N+\delta_{ij}\hbar)}    & a_2=0\,, \\
   \f1{\pi z} & a_2\neq 0\,.
 \end{array}
 \right.
$$
$$
\rho_{ij}^0=\f1{\pi l\hbar}+\f1{\pi lu_{ji}}\,.
 $$

For elliptic Baxter $R$-matrix there exists another trigonometric and rational limits \cite{GZ,StK,AKZ,Sm}. This construction
can be generalized to $\SLN$ elliptic matrix
This approach
can be applied in our case as well. We will not  develop this issue here.

\section{Dynamical $R$-Matrices and  Integrable systems}
\setcounter{equation}{0}

\subsection{IRF models}
Following \cite{Fe1} we construct the Boltzmann  weights of the interaction-round-the-face models
starting with the
quantum $R$-matrices described above. Let $\mu\in\gh^*$ be a weight of  ($\mu\in\gh^*$) of the
vector representation of $\sln$ in $V$. In other words we have $N$ weights
\beq{wm}
P_V=\{\mu_j=\f1{N}(-1,\ldots,-1,N-1,-1,\ldots,-1)\,,~~(j=1,\ldots,N)\}\,,
\eq
 where $N-1$ stays on the $j$ place. Let
$V[\mu_j]$ be the corresponding component of the space $V$, and $E[\mu_j]\,:\,V\to V[\mu_j]$ is a projection.
In our case all $V[\mu_j]$ are one-dimensional.

Define the
local states $a,b,c,d\in\gh^*$  of the IRF model
$b-a=\mu_4$, $c-b=\mu_3$, $d-c=\mu_2$, $d-a=\mu_1$,
where  all weights from $P_V$ (\ref{wm}), satisfy the equality
$\mu_1+\mu_2=\mu_3+\mu_4$.

Define the map
$W(a,b,c,d)\,:\,V_{\mu_1}\otimes V_{\mu_2}\to V_{\mu_4}\otimes V_{\mu_3}$
by means of the $R$-matrix
 \beq{Bowe}
W(a,b,c,d,z,\bfu)=E[c-b]\otimes E[b-a]R(\bfu+\hbar a+\hbar c,z)|_{V[d-a]\otimes V[c-d]}\,.
 \eq
 In fact
$W(a+\ti\bfu,b+\ti\bfu,c+\ti\bfu,d+\ti\bfu,\bfu,z-2\hbar\ti\bfu)$ is independent on $\ti\bfu$. In this way we can
 define \emph{\underline{the Boltzmann weights}} of the IRF model as $W(a,b,c,d,z)=W(a,b,c,d,0,z)$.
The partition function of the IRF model takes the form
 $$
Z=\sum_{lattice}W(a_{ij},a_{i,j+1},a_{i-1,j+1},a_{i-1,j-1})\,,
 $$

\unitlength 1mm 
\bigskip
\begin{center}
\unitlength 0.5mm 
\linethickness{0.4pt}
\ifx\plotpoint\undefined\newsavebox{\plotpoint}\fi 
\begin{picture}(79.5,88.25)(0,0)
\put(22.75,30.25){\framebox(60,51.75)[lb]{}}
\put(14.5,86){a}
\put(88,87.75){b}
\put(87,29.25){c}
\put(15.5,30.5){d}
\put(47.75,88.25){$\mu_4$}
\put(89.5,58.75){$\mu_3$}
\put(45.25,24.5){$\mu_2$}
\put(15.5,59.25){$\mu_1$}
\end{picture}

\end{center}

If $R$ satisfies quantum dynamical Yang-Baxter (QDYB) equation (\ref{z1}), then $W$ obeys

\underline{ Star-Triangle relations} \cite{Fe}
 $$
\sum_gW^{12}(b,c,d,g,z_{12})W^{13}(a,b,g,f,z_{13})W^{23}(f,g,d,c,z_{23)}=
 $$
 $$
\sum_gW^{23}(a,b,c,g,z_{23})W^{13}(g,c,d,e,z_{13})W^{12}(a,g,e,f,z_{12})\,.
 $$
on $V[f-a]\otimes V[e-f]\otimes V[d-e]$ providing the integrability of the model.

\subsection{Elliptic Quantum Groups}

Let $R$ be a solution of QDYB. Define the quantum Lax operator $L(\hat\bfv,\bfu,\hat S,z)$ as a map of $\ti\gh_0\times\mC$ to
Aut$(V)$. Here $\hat S$ are generators of $\SLN$ acting
 on the module $V$, and $[\hat v_j,u_k]=\hbar\de_{jk}$. The Lax operator satisfies  the equation
 \beq{rll}
R^{12}(z-w,\bfu+\hbar e^3)L^{1}(\bfu-\hbar e^2,z)L^{2}(\bfu+\hbar e^1,w)=
L^{2}(\bfu-\hbar e^1,w)L^{1}(\bfu+\hbar e^2)R^{12}(\bfu-\hbar e^3\bfu,z-w)\,,
 \eq
 where $L^{1}=L\otimes Id$, $L^{2}=Id\otimes L$.

Assume that  $L(\hat\bfv,\bfu,z,\hat S)$  satisfies the quasi-periodicity conditions (\ref{qp})
 and the weight zero condition (\ref{z}).
 The relation (\ref{rll}) defines the quadratic algebra with respect to $\bfu$ and $\hat S$.
 This algebra is the Felder elliptic quantum group \cite{Fe} for the trivial bundles and $R$ (\ref{z3}).
 In the case $R$ (\ref{z2}) we come to the Sklyanin-Feigin-Odesski algebras \cite{Skl1,FO,CLOZ2,KLO}.

 In the classical limit $L=L(\hat\bfv,\bfu,\hat S,z)$ becomes the classical Lax operator for interacting tops
 described in section 3.1.
 The angular momentum variable $S$ belongs to the coadjoint SL$(p,\mC)$ orbit corresponding after
 quantization to the representation $V$.
  \mo{Here $V$ is a representation of $\SLN$, but not SL$(p,\mC)$ ?
 Or we deal here with restriction of irreducible  $\SLN$-module $V$ to reducible  SL$(p,\mC)$ modules.}
In this way we obtain the quadratic Poisson algebras
 $$
 \{L^1(\bfu,S),L^2(\bfu,S)\}=[r(\bfu),L^1(\bfu,S),L^2(\bfu,S)]
$$
defining the Poisson structure on the phase space of  interacting tops.



\section{Appendix: Elliptic Functions }
\setcounter{equation}{0}
\def\theequation{A.\arabic{equation}}
\subsection{Basic Definitions and Properties}
Let $q=\exp (2\pi i\tau)$, where $\tau$ is the modular
parameter of the elliptic curve $E_\tau$.
The basic element is the theta  function:
\beq{A.1a}
 \vth(z|\tau)=q^{\frac {1}{8}}\sum_{n\in {\bf Z}}(-1)^n\bfe(\oh
n(n+1)\tau+nz)
  \eq
 $$
=q^{\frac{1}{8}}e^{-\frac{i\pi}{4}} (e^{i\pi z}-e^{-i\pi z})
\prod_{n=1}^\infty(1-q^n)(1-q^ne^{2i\pi z})(1-q^ne^{-2i\pi z})\,.
~~(\bfe(x)=\exp 2\pi\imath x)
  $$
\bigskip

\underline{{\it The  Eisenstein functions}}
 \beq{A.1}
E_1(z|\tau)=\p_z\log\vth(z|\tau), ~~E_1(z|\tau)\sim\f1{z}-2\eta_1z,
 \eq where

 \beq{A.6} \eta_1(\tau)=\frac{3}{\pi^2}
\sum_{m=-\infty}^{\infty}\sum_{n=-\infty}^{\infty '} \frac{1}{(m\tau+n)^2}=\frac{24}{2\pi i}\frac{\eta'(\tau)}{\eta(\tau)}\,
\eq and $ \eta(\tau)=q^{\frac{1}{24}}\prod_{n>0}(1-q^n)\, $ is the Dedekind function.

 \beq{A.2}
  E_2(z|\tau)=-\p_zE_1(z|\tau)= \p_z^2\log\vth(z|\tau),
~~E_2(z|\tau)\sim\f1{z^2}+2\eta_1\,.
\eq
The higher Eisenstein functions
 \beq{A.2a}
  E_j(z)=\frac{(-1)^j}{(j-1)!}\p^{(j-2)}E_2(z)\,,~~(j>2)\,.
 \eq

\underline{{\it Relation to the Weierstrass functions}}
\beq{a100}
\zeta(z,\tau)=E_1(z,\tau)+2\eta_1(\tau)z\,,
\eq
 \beq{a101}
 \wp(z,\tau)=E_2(z,\tau)-2\eta_1(\tau)\,.
  \eq

The next important function is
\beq{A.3}
 \phi(u,z)= \frac {\vth(u+z)\vth'(0)} {\vth(u)\vth(z)}\,.
  \eq
   \beq{A.300}
\phi(u,z)=\phi(z,u)\,,~~\phi(-u,-z)=-\phi(u,z)\,.
 \eq
 It has a pole at $z=0$ and
  \beq{A.3a}
\phi(u,z)=\frac{1}{z}+E_1(u)+\frac{z}{2}(E_1^2(u)-\wp(u))+\ldots\,.
  \eq
 \emph{ \underline{Trigonometric limit }}  for $\phi(u,z)$ follows from (\ref{A.1a})
  \beq{A.3z}
\lim_{\Im m\tau\to+\infty}\phi(u,z)=\frac{\sin\pi (z+u)}{\sin\pi z\sin\pi u}\,.
  \eq

Let $f(u,z)=\p_u\phi(u,z)$. Then
 \beq{A3b}
 f(u,z)=\phi(u,z) (E_1(u+z)-E_1(u))\,.
  \eq

\underline{{\it Heat equation}}  \beq{A.4b} \p_\tau\phi(u,w)-\f1{2\pi i}\p_u\p_w\phi(u,w)=0\,.  \eq

\underline{{\it Quasi-periodicity}}

 \beq{A.11}
  \vth(z+1)=-\vth(z)\,,~~~\vth(z+\tau)=-q^{-\oh}e^{-2\pi
iz}\vth(z)\,,  \eq  \beq{A.12} E_1(z+1)=E_1(z)\,,~~~E_1(z+\tau)=E_1(z)-2\pi i\,,
 \eq
  \beq{A.13}
E_2(z+1)=E_2(z)\,,~~~E_2(z+\tau)=E_2(z)\,,  \eq  \beq{A.14} \phi(u,z+1)=\phi(u,z)\,,~~~\phi(u,z+\tau)=e^{-2\pi \imath
u}\phi(u,z)\,.  \eq  \beq{A.15} f(u,z+1)=f(u,z)\,,~~~f(u,z+\tau)=e^{-2\pi \imath u}f(u,z)-2\pi\imath\phi(u,z)\,.
  \eq

\underline{{\it  The Fay three-section formula:}}

 \beq{ad3}
\phi(u_1,z_1)\phi(u_2,z_2)-\phi(u_1+u_2,z_1)\phi(u_2,z_2-z_1)- \phi(u_1+u_2,z_2)\phi(u_1,z_1-z_2)=0\,.  \eq Particular cases
of this formula is the  Calogero functional equation

 \beq{ad2}
\phi(u,z)\p_v\phi(v,z)-\phi(v,z)\p_u\phi(u,z)=(E_2(u)-E_2(v))\phi(u+v,z)\,,
 \eq

 \beq{ad21} \phi(u,z)\phi(-u,z)=E_2(z)-E_2(u).  \eq

Another important relation is  \beq{ir} \phi(v,z-w)\phi(u_1-v,z)\phi(u_2+v,w) -\phi(u_1-u_2-v,z-w)\phi(u_2+v,z)\phi(u_1-v,w)=
 \eq
 $$
\phi(u_1,z)\phi(u_2,w)f(u_1,u_2,v)\,,
 $$
where  \beq{ir1} f(u_1,u_2,v)=E_1(v)-E_1(u_1-u_2-v)+E_1(u_1-v)-E_1(u_2+v)\,.  \eq
Taking limit
$u_2\rightarrow 0$ in (\ref{ir}) we obtain:

 \beq{ir21} \phi(v,z-w)\phi(u_1-v,z)\phi(v,w)
-\phi(u_1-v,z-w)\phi(v,z)\phi(u_1-v,w)=  \eq
 $$
\phi(u_1,z)(E_2(v)-E_2(u_1-v)),
 $$
which is equivalent to (\ref{ad2}) due to (\ref{A3b}).

\bigskip
\underline{{\it Theta functions with characteristics:}}\\
For $a, b \in \Bbb Q$ by definition:
 $$\theta{\left[\begin{array}{c}
a\\
b
\end{array}
\right]}(z , \tau ) =\sum_{j\in \Bbb Z} {\bf
e}\left((j+a)^2\frac\tau2+(j+a)(z+b)\right)\,.
 $$
In particular, the function $\vth$ (\ref{A.1a}) is a theta function with characteristics:  \beq{A.29}
\vartheta(x,\tau)=\theta\left[
\begin{array}{c}
1/2\\
1/2
\end{array}\right](x,\tau)\,.
 \eq Properties:
 $$
\theta{\left[\begin{array}{c}
a\\
b
\end{array}
\right]}(z+1 , \tau )={\bf e}(a) \theta{\left[\begin{array}{c}
a\\
b
\end{array}
\right]}(z  , \tau )\,,
 $$
 $$
\theta{\left[\begin{array}{c}
a\\
b
\end{array}
\right]}(z+a'\tau , \tau ) ={\bf e}\left(-{a'}^2\frac\tau2
-a'(z+b)\right) \theta{\left[\begin{array}{c}
a+a'\\
b
\end{array}
\right]}(z , \tau )\,,
 $$
 $$\theta{\left[\begin{array}{c}
a+j\\
b
\end{array}
\right]}(z , \tau )= \theta{\left[\begin{array}{c}
a\\
b
\end{array}
\right]}(z , \tau ),\quad j\in \Bbb Z\,.
 $$

\subsection{Lie Group $\GLN$ and Elliptic Functions}

Introduce the notation (see \cite{FFZ})
 $$
{\bf e}_N(z)=\exp (\frac{2\pi i}{N} z)
 $$
 and two matrices
 \beq{q}
 Q=\di({\bf e}_N(1),\ldots,{\bf e}_N(m),\ldots,1)
  \eq
 \beq{la}
  \La= \left(\begin{array}{ccccc}
0&1&0&\cdots&0\\
0&0&1&\cdots&0\\
\vdots&\vdots&\ddots&\ddots&\vdots\\
0&0&0&\cdots&1\\
1&0&0&\cdots&0
\end{array}\right)\,.
 \eq
 We have $Q\La=\exp -(\frac{2\pi i}{N})\La Q$.
 Let
 \beq{B.10}
  \Gamma_N=\mZ^{(2)}_N=(\mZ/N\mZ\oplus\mZ/N\mZ)\,,~~\ti{\Gamma}_N=\ti{\mZ}^{(2)}_N=
\mZ^{(2)}_N\setminus(0,0)
\eq
 be the two-dimensional lattice of order $N^2$ and $N^2-1$ correspondingly. The matrices
$Q^{a_1}\La^{a_2}$, $a=(a_1,a_2)\in\mZ^{(2)}_N$ generate a basis in the group $\GLN$, while $Q^{\al_1}\La^{\al_2}$,
$\al=(\al_1,\al_2)\in\ti{\mZ}^{(2)}_N$ generate a basis in the Lie algebra $\gln$.
Consider the projective representation of
$\mZ^{(2)}_N$ in $\GLN$
 \beq{B.11}
  a\to T_{a}= \frac{N}{2\pi i}\bfe_N(\frac{a_1a_2}{2})Q^{a_1}\La^{a_2}\,,
  \eq
   \beq{AA3a}
T_aT_b=\ka_{a,b}T_{a+b}\,,\ \ \ka_{a,b}=\frac{N}{2\pi i}\bfe_N(-\frac{a\times b}{2}),\ ~~ (a\times b=a_1b_2-a_2b_1)
 \eq
 Let us mention some simple properties of $\ka$:

 \beq{kap} \ka_{a,b}\ka_{b,a}=\left(\frac{N}{2\pi i}\right)^2,\ \
\ka_{a,c}\ka_{b,c}=\frac{N}{2\pi i}\ka_{a+b,c},\ \ \ka_{a,a}=\frac{N}{2\pi i}.  \eq

Note that $\ka_{a,b}$ can be interpreted as a non-trivial two-cocycle in $H^2(\mZ^{(2)}_N,\mZ_{2N})$. It follows from
(\ref{AA3a}) that  \beq{AA3b} [T_{\al},T_{\be}]=\bfC(\al,\be)T_{\al+\be}\,,  \eq where
$\bfC(\al,\be)=\frac{N}{\pi}\sin\frac{\pi}{N}(\al\times \be)$ are the structure constants of $\gln$.

\

\underline{\it Deformed Elliptic Functions}

 \beq{90}
\vf_a(\eta,z)=\bfe_N(a_2z)\phi(\om_a+\eta,z)\,,~\om_a=\frac{a_1+a_2\tau}{N}\,,~ a\in\mZ^{(2)}_N\,,~\eta\in\Si_\tau\,,
 \eq
 and
  \beq{kfi}
 \varphi^m_\be(\bfu,z)=\bfe\,(\lan\ka,\be\ran z)\phi(\lan \bfu+\ka\tau,\be\ran+\frac{m}l,z)\,,
\eq
 \emph{ \underline{Trigonometric limit }}  for $ \vf_a(\eta,z)$ (see (\ref{A.3z})
 \beq{A.3zz}
 \lim_{\Im m\tau\to+\infty}\vf_a(\eta,z)=\left\{
 \begin{array}{cc}
\cot\pi z+\cot\pi(\eta+\frac{a_1}N)    & a_2=0\,, \\
   \bfe((\frac{a_2}N+1)z)\sin^{-1}\pi z & a_2\neq 0\,.
 \end{array}
 \right.
 \eq
It follows from (\ref{A.14}) that $\vf_a(z,\eta)$ is well defined on
$\mZ^{(2)}_N$:
 \beq{91}
 \vf_{a+c}(\eta,z)=\vf_{a}(\eta,z)\,,~{\rm
for}~c_{1,2}\in\mZ~{\rm mod} \,N\,
 \eq
 \beq{92}
  \vf_a(\eta,z+1)=\bfe_N(a_2)\vf_a(\eta,z)\,,~~
\vf_a(\eta,z+\tau)=\bfe_N(-a_1-N\eta)\vf_a(\eta,z)\,.
\eq
 \beq{qpu}
  \vf_a(\eta+1,z)=\vf_a(\eta,z)\,,~~
\vf_a(\eta+\tau,z)=\bfe(-z)\vf_a(\eta,z)\,.
\eq

For $ \varphi^m_\be(\bfu,z)$ (\ref{kfi}) we have
\beq{A.14a}
\varphi_\be^{m}(\bfu,z+1)=\bfe\,(\lan\ka,\be\ran \varphi_\be^{m}(\bfu,z)\,,~~~
\varphi_\be^{m}(\bfu,z+\tau)=\bfe(- \lan\bfu,\be\ran-\frac{m}l)\varphi_\be^{m}(\bfu,z)\,.
\eq
The following formulae can be proved directly by checking the
structure of poles and quasi-periodic properties:

 \beq{a93}
 \sum\limits_{a\in \mZ^{(2)}_N} E_2(z+\om_a)=N^2E_{2}(Nz)
 \eq

By analogy with (\ref{ad3}) and (\ref{ir}-\ref{ir21}) we have:
 \beq{a94}
\begin{array}{c}
\vf_{a+b}(z-w,u+\om_a+\om_b)\vf_{-b}(z,v-\om_b)+
\vf_{a}(z,u+v+\om_a)\vf_{-a-b}(w,-u-\om_a-\om_b)
\\
=\vf_a(z-w,u+v+\om_a)\vf_{-b}(w,v-\om_b)
\end{array}
 \eq

 \beq{a95}
\begin{array}{c}
\vf_{a+b+c}(z-w,u+\om_a+\om_b+\om_c)\vf_{-b-c}(z,v-\om_b-\om_c)\vf_c(w,u+\om_c)-
\\
\vf_{a-c}(z-w,v+\om_a-\om_c)\vf_{c}(z,u+\om_c)\vf_{-b-c}(w,v-\om_b-\om_c)=
\end{array}
 \eq
 $$
=
\left\{
\begin{array}{l}
\vf_a(z,\om_a+u+v)\vf_{-a-b}(w,-\om_a-\om_b)
(E_1(u+\om_a+\om_b+\om_c)-E_1(v+\om_a-\om_c)+
\\
E_1(v-\om_b-\om_c)-E_1(u+\om_c)),\ \ \hbox{if}\ a+b\neq 0\
\hbox{mod}\ \mZ^{(2)}_N,
\\
\
\\
\vf_a(z,\om_a+u+v)(E_2(u+\om_c)-E_2(v+\om_a-\om_c)),\ \ \hbox{if}\
a+b=0\ \hbox{mod}\ \mZ^{(2)}_N.
\end{array}
\right.
 $$



\small{

\end{document}